\documentclass{aastex6}
\usepackage{mathtools}
\usepackage{amsmath}
\usepackage{float}
\usepackage{longtable}
\usepackage{soul}

\fullcollaborationName{The Friends of AASTeX Collaboration}

\begin{document}

\title{Precision Orbit of $\delta$ Delphini and Prospects for Astrometric Detection of Exoplanets}

\author{Tyler Gardner\altaffilmark{1}, John D. Monnier\altaffilmark{1}, Francis C. Fekel\altaffilmark{2}, Mike Williamson\altaffilmark{2}, Douglas K. Duncan\altaffilmark{3}, Timothy R. White\altaffilmark{10}, Michael Ireland\altaffilmark{13}, Fred C. Adams\altaffilmark{12}, Travis Barman\altaffilmark{16}, Fabien Baron\altaffilmark{15}, Theo ten Brummelaar\altaffilmark{14}, Xiao Che\altaffilmark{1}, Daniel Huber\altaffilmark{7}\altaffilmark{8}\altaffilmark{9}, Stefan Kraus\altaffilmark{5}, Rachael M. Roettenbacher\altaffilmark{4}, Gail Schaefer\altaffilmark{14}, Judit Sturmann\altaffilmark{14}, Laszlo Sturmann\altaffilmark{14}, Samuel J. Swihart\altaffilmark{6}, Ming Zhao\altaffilmark{11}}

\altaffiltext{1}{Astronomy Department, University of Michigan, Ann Arbor, MI 48109, USA}
\altaffiltext{2}{Center of Excellence in Information Systems, Tennessee State University, Nashville, TN 37209, USA}
\altaffiltext{3}{Dept. of Astrophysical and Planetary Sciences, Univ. of Colorado, Boulder,
Colorado 80309, USA}
\altaffiltext{4}{Department of Astronomy, Stockholm University, SE-106 91 Stockholm, Sweden}
\altaffiltext{5}{University of Exeter, School of Physics, Astrophysics Group, Stocker Road, Exeter EX4 4QL, UK}
\altaffiltext{6}{Department of Physics and Astronomy, Michigan State University, East Lansing, MI 48824, USA}
\altaffiltext{7}{Institute for Astronomy, University of Hawai`i, 2680 Woodlawn Drive, Honolulu, HI 96822, USA}
\altaffiltext{8}{Sydney Institute for Astronomy (SIfA), School of Physics, University of Sydney, NSW 2006, Australia}
\altaffiltext{9}{SETI Institute, 189 Bernardo Avenue, Mountain View, CA 94043, USA}
\altaffiltext{10}{Stellar Astrophysics Centre, Department of Physics and Astronomy, Aarhus University, Ny Munkegade 120, DK-8000 Aarhus C, Denmark}
\altaffiltext{11}{Department of Astronomy \& Astrophysics, The Pennsylvania State University, 525 Davey Lab, University Park, PA 16802}
\altaffiltext{12}{Physics Department, University of Michigan, Ann Arbor, MI 48109, USA}
\altaffiltext{13}{Research School of Astronomy \& Astrophysics, Australian National University, Canberra
ACT 2611, Australia}
\altaffiltext{14}{The CHARA Array of Georgia State University, Mount Wilson Observatory, Mount Wilson, CA 91203, USA}
\altaffiltext{15}{Department of Physics and Astronomy, Georgia State University, Atlanta, GA, USA}
\altaffiltext{16}{Lunar and Planetary Laboratory, University of Arizona, Tucson AZ 85721 USA}

\begin{abstract}
Combining visual and spectroscopic orbits of binary stars leads to a determination of the full 3D orbit, individual masses, and distance to the system. We present a full analysis of the evolved binary system $\delta$ Delphini using astrometric data from the MIRC and PAVO instruments on the CHARA long-baseline interferometer, 97 new spectra from the Fairborn Observatory, and 87 unpublished spectra from Lick Observatory. We determine the full set of orbital elements for $\delta$ Del, along with masses of $1.78 \pm 0.07$ $M_{\odot}$ and $1.62 \pm 0.07$ $M_{\odot}$ for each component, and a distance of $63.61 \pm 0.89$ pc. These results are important in two contexts: for testing stellar evolution models and defining the detection capabilities for future planet searches. We find that the evolutionary state of this system is puzzling, as our measured flux ratios, radii, and masses imply a $\sim$ 200 Myr age difference between the components using standard stellar evolution models. Possible explanations for this age discrepancy include mass transfer scenarios with a now ejected tertiary companion. For individual measurements taken over a span of 2 years we achieve $<10$ $\mu$-arcsecond precision on differential position with 10-minute observations. The high precision of our astrometric orbit suggests that exoplanet detection capabilities are within reach of MIRC at CHARA. We compute exoplanet detection limits around $\delta$ Del, and conclude that if this precision is extended to wider systems we should be able to detect most exoplanets $>2$ M$_{J}$ on orbits $>0.75$ AU around individual components of hot binary stars via differential astrometry. 

\end{abstract}


\section{Introduction} 
\label{sec:intro}
Binary systems provide a unique opportunity for studying the physical properties of stars. Combining spectroscopic and astrometric studies of binary stars allows one to determine the full 3D orbit of the system and obtain fundamental properties such as masses and distance. Systems for which both double-lined spectroscopic and visual orbits can be obtained are therefore valuable systems for the testing of stellar evolution models. Long-baseline interferometry provides the capability for resolving sub-arcsecond binary systems in order to obtain visual orbits of systems that would otherwise only be resolved through spectroscopic studies. \citet{bonneau2014} give a thorough overview on how interferometric studies are combined with spectroscopy to determine the physical properties of both components in a binary system. In this paper we use the Michigan Infra-Red Combiner (MIRC) on the Center for High Angular Resolution Astronomy (CHARA) Array long-baseline interferometer to obtain a precise visual orbit of the close binary system $\delta$ Delphini (HR 7928, HD 197461). With the visual orbit we achieve $<10$ $\mu$-arcsecond precision, maintained over 2 years, on many of the individual measurements of differential position.

Because of the short-period variations in its light curve, $\delta$ Del was first classified as a $\delta$ Scuti variable by \citet{Eggen1956}. \citet{Struve1957} confirmed this variable star classification through a spectroscopic study of $\delta$ Del. Neither of these studies detected the binarity of the system. As part of
the Reports of Observatories, 1965-1966, published in The Astronomical Journal, 
Whitford reported in the Lick Observatory yearly summary that G. Preston
had discovered $\delta$~Del to be a double lined spectroscopic binary with a preliminary period of 40 days. The high eccentricity of the system produced double lines that are only visible for about three of the 40 days, which is the reason why previous observers had not discovered the binarity of the system. From an undergraduate thesis by Duncan in 1973, \citet{Duncan1979} reported the results of the first comprehensive study of $\delta$ Del as a binary system. Using Lick Observatory spectra, they obtained radial velocities (RVs) from which they determined a binary orbit with a period of $40.580$ days and an high eccentricity of 0.7. They also found that both the primary (more massive) and secondary components show $\delta$ Scuti pulsations with dominant periods of 0.158 and 0.134 days, respectively. They concluded that the components were nearly equal in luminosity and temperature but determined a mass ratio of $\sim1.2$. This made it impossible for Duncan and Preston to find locations in the Hertzsprung-Russell (HR) diagram that satisfied the constraints of mass ratio, luminosity, and the stars being the same age. In this paper we combine our astrometric data from CHARA, radial velocities from 97 new spectra obtained at Fairborn Observatory, and the unpublished radial velocities from the 87 Lick Observatory spectra measured by \citet{Duncan1979} to obtain a 3D orbit of $\delta$~Del. We also reassess the age and other properties of the system using stellar evolution models. 

Along with our orbital study of $\delta$ Del, we use the $<10$ $\mu$-as precision demonstrated on this system to explore the feasibility of detecting exoplanets around stars in a close binary system using MIRC at CHARA. A Jupiter mass planet at a separation of $1$ AU imparts about a $10$ $\mu$-as wobble on a solar mass host star at the distance of $\delta$ Del. Thus, with the precision of MIRC we should be able to detect this wobble on a single component of a close binary system. Astrometric orbits of planets are desirable since they unveil important orbital parameters such as the inclination of the orbit and true mass. Unlike radial velocity or transit methods, astrometric detection is favorable for planets that have wider orbits. On the other hand, astrometry is sensitive to planets on somewhat tighter orbits than direct imaging surveys. Although this regime is comparable to that explored through microlensing techniques, astrometry has the advantage of repeat observations. Moreover, planet detection via differential astrometry with the use of long-baseline interferometry favors A and B-type binary stars. This is a regime that is very difficult to explore with radial velocity surveys because hot stars typically have  weak and broad spectral lines. Transit surveys are also biased against these stars since stellar pulsations and variability mask transit signals. Historically, the exoplanet field has been riddled with false claims of detection via astrometry (see \citet{Muterspaugh2010} for a brief overview). However, as instrumental precision continues to improve, astrometric detection of exoplanets is finally becoming feasible. By the end of its nominal five year mission, Gaia is expected to reveal many new astrometric detections of giant exoplanets around mostly lower-mass stars \citep{Perryman2014,sahlmann2014,casertano2008,sozzetti2014}. From the ground, long-baseline interferometry is a promising method for detecting exoplanets around intermediate mass stars in close binary systems. The Palomar High-precision Astrometric Search for Exoplanet Systems (PHASES) recently used long-baseline interferometry that led to the announcement of  six substellar candidates to the individual components of binaries \citep{Muterspaugh2010}. In this paper we show that the MIRC instrument at CHARA has achieved the precision needed for exoplanet detection around single stars in close (sub-arcsecond) binary systems.

This paper is organized as follows. Section \ref{sec:observations} describes our observations and the subsequent data reduction. Section \ref{sec:orbit} then outlines our orbit fitting techniques, and Section \ref{sec:fitting_results} presents the best fit orbital and physical  parameters for the $\delta$ Del binary system. In Section \ref{sec:hr}, we use stellar evolution considerations to interpret the unusual positions of the $\delta$ Del components in the H-R diagram. The paper concludes, in Section \ref{sec:exoplanets}, with a discussion of the corresponding limits on future exoplanet detections.
\clearpage

\section{Observations and Data Reduction} 
\label{sec:observations}
\subsection{Interferometry}
\label{sec:interferometry}
Interferometric data for $\delta$ Del were collected in $H$-band on eleven nights from 2011 July 15 to 2013 July 14 with MIRC at the CHARA Array. The CHARA Array is an optical/near IR interferometer with the longest baselines of any interferometer of its type in the world \citep{tenbrummelaar2005}. MIRC combines all six telescopes available at CHARA with baselines up to 330 meters. The instrument is described in detail by \citet{Monnier2006}. Additionally, R-band data were recently obtained with the Precision Astronomical Visible Observations (PAVO) instrument in 2017 June 14-17. PAVO is a visible light beam combiner on the CHARA array which is predominantly used for two-telescope observations. The PAVO instrument and data reduction techniques are described further in \citet{ireland2008}. Observational details and calibrators used for MIRC are displayed in Tables \ref{mirctbl-1} and \ref{mirctbl-2}, while those for PAVO are given in Tables \ref{pavotbl-1} and \ref{pavotbl-2}. The angular diameters for the calibrators in Table \ref{pavotbl-2} were obtained from the $V-K$ surface brightness relation of \citet{kervella2004}.

\begin{table} [H]
\begin{center}
\caption{Log of MIRC interferometric observations.\label{mirctbl-1}}
\begin{tabular}{llcc}
\tableline
\tableline
\textsc{UT} date & Baseline & No. of 10-sec averages & Calibrators\tablenotemark{2} \\
\tableline
2011 Jul 15 & S2E1W1W2E2 & 168 & a \\
2011 Jul 17 & S1S2E1W1W2 & 80 & b \\
2012 June 10 & W1W2E2 & 48 & c \\
2012 June 12 & S1S2W1W2E2 & 160 & d \\
2012 June 15 & S1S2E1W1W2E2 & 120 & e \\
2012 June 16 & S1S2W1W2 & 48 & f \\
2012 June 20 & S1S2W1W2E2 & 80 & g \\
2012 Sep 19 & S1S2E1W1W2E2 & 120 & h \\
2012 Sep 20 & S1S2W1W2E2 & 80 & i \\
2013 Jul 13 & S2W1W2 & 24 & a \\
2013 Jul 14 & S1S2E1W1W2E2 & 120 & e \\
\tableline
\end{tabular}
\tablenotetext{2}{Refer to Table~\ref{mirctbl-2} for details of the calibrators used.}
\end{center}
\end{table}

\begin{table} [H]
\begin{center}
\caption{Calibrators used for MIRC interferometric observations.\label{mirctbl-2}}
\begin{tabular}{lcccccc}
\tableline
\tableline
HD & Sp. type & $H$ (mag) & $\theta_\mathrm{UD}$ (mas) & Source for UD & ID \\
\tableline
205776 & K2III & 4.138 & $0.79 \pm 0.055$ & \citet{chelli2016} & a \\
886 & B2IV & 3.43 & $0.41 \pm 0.03$ & \citet{barnes1978} & b \\
135742 & B8Vn & 2.8 & $0.645 \pm 0.045$ & \citet{chelli2016} & c \\
165777 & A5V & 3.426 & $0.68 \pm 0.06$ & \citet{chelli2016} & d \\
187691 & F8V & 3.863 & $0.7 \pm 0.04$ & \citet{chelli2016} & e\\
185395 & F3+V & 3.716 & $0.726 \pm 0.014$ & \citet{white2013} & f \\
161868 & A1VnkA0mA0 & 3.64 & $0.571 \pm 0.04$ & \citet{chelli2016} & g \\
6920 & 	F8V & 4.493 & $0.539 \pm  0.037$ & \citet{chelli2016} & h \\
195810 & B6III & 4.55 & $0.35 \pm 0.05$ & \citet{barnes1978} & i \\
\tableline
\end{tabular}
\end{center}
\end{table}

\begin{table} [H]
\begin{center}
\caption{Log of PAVO interferometric observations.\label{pavotbl-1}}
\begin{tabular}{lccc}
\tableline
\tableline
\textsc{UT} date & Baseline\tablenotemark{1} & No. of scans & Calibrators\tablenotemark{2} \\
\tableline
2017 June 14 & E2W1 & 3 & cd \\
2017 June 15 & E1W2 & 3 & cd \\
2017 June 17 & E2W2 & 5 & cd \\
2017 June 18 & S1W2 & 3 & abcd \\
2017 June 19 & W1W2 & 3 & ac \\
\tableline
\end{tabular}
\tablenotetext{1}{The baselines used have the following lengths:
W1W2, 107.92\,m; E2W2, 156.27\,m; S1W2, 210.97\,m; E1W2, 221.82\,m; E2W1, 251.33\,m.}
\tablenotetext{2}{Refer to Table~\ref{pavotbl-2} for details of the calibrators used.}
\end{center}
\end{table}

\begin{table} [H]
\begin{center}
\caption{Calibrators used for PAVO interferometric observations.\label{pavotbl-2}}
\begin{tabular}{lccrccc}
\tableline
\tableline
HD & Sp. type & $V$ & $V-K$ & $A_V$ & $\theta_{V-K}$ & ID \\
\tableline
195943 & A3IVs & 5.380 &    0.138 & 0.089 & 0.299 & a \\
196775 & B3V   & 5.960 & $-$0.473 & 0.261 & 0.153 & b \\
196821 & A0III & 6.075 &    0.034 & 0.000 & 0.204 & c \\
201616 & A2Va  & 6.057 &    0.117 & 0.000 & 0.218 & d \\
\tableline
\end{tabular}
\end{center}
\end{table}

\begin{figure}[H]
\centering
\includegraphics[width=\linewidth]{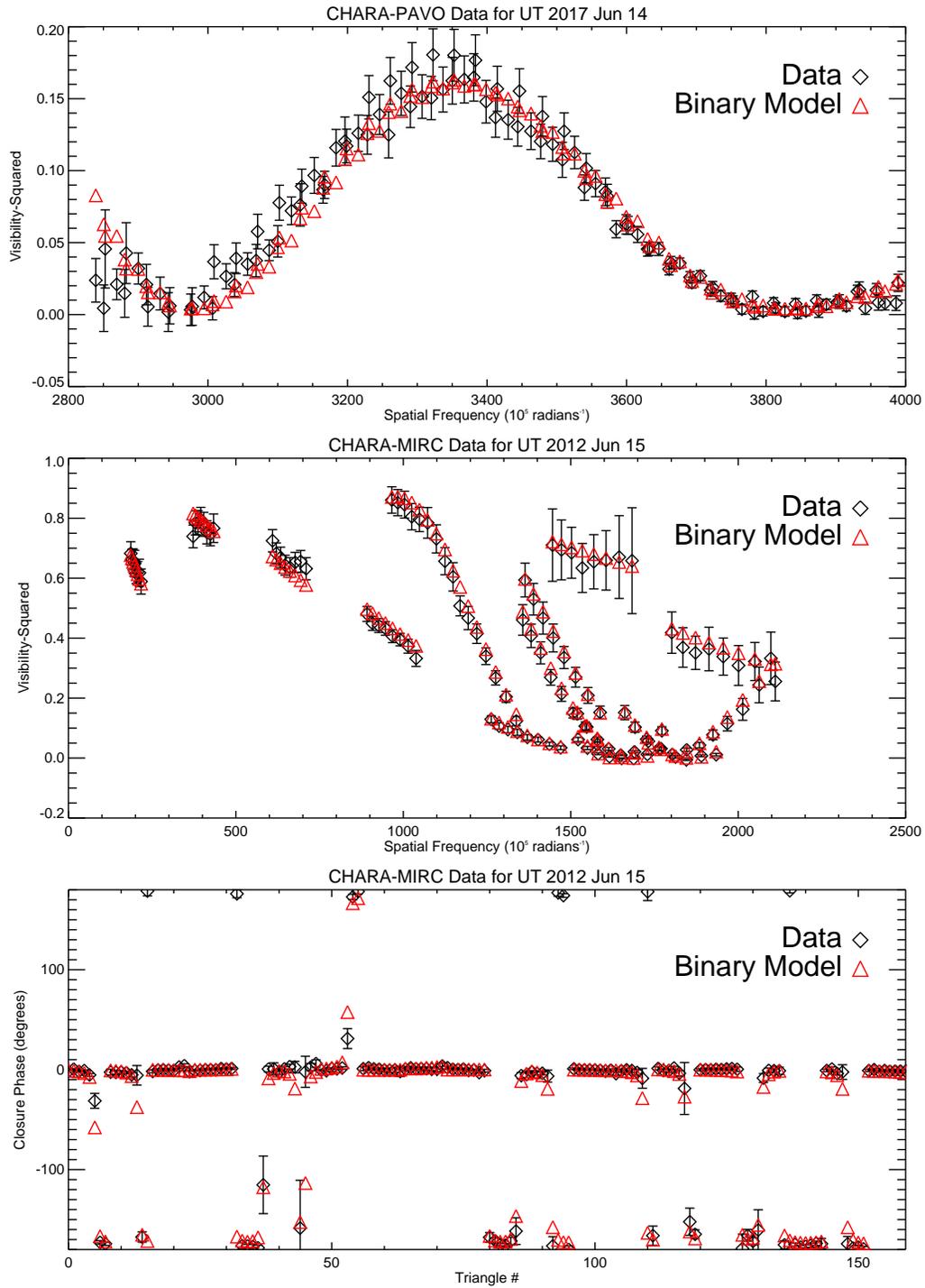}
\caption{Squared visibilities are plotted along with the best fit binary model on a single night for the PAVO and MIRC interferometric data. For the MIRC data we also plot closure phases for a single night. The "Triangle \#" is a combination of time, geometry (which closing triangle), and wavelength.}
\label{visplot}
\end{figure}

We used the MIRC combiner to measure visibilities and closure phases of $\delta$ Del. Amplitude calibration was performed through use of a beamsplitter following spatial filtering. Observations of reference calibrators are made throughout the night to correct for time-variable factors such as atmospheric coherence time, vibrations, differential dispersion, and birefringence in the beam train. Using the standard data pipeline as described in earlier MIRC papers \citep[e.g.][]{monnier2012}, we produce a calibrated OI-FITS file \citep{pauls2005} for each night (available upon request).  For each night we fit a binary model with the following free parameters:  Uniform Disk (UD) diameter of component 1, UD diameter of component 2, H band flux ratio of component 1 over component 2, angular separation, position angle (PA) of vector pointing from component 1 to 2 (east of north).  To estimate errors we derive a $\chi^2$ surface for a grid in relative Right Ascension (RA) and Declination (Dec) and find the $1-\sigma$ confidence contour (approximated by an ``error ellipse'' with a major axis, minor axis and PA of major axis) -- for this, we made a simple assumption that the errors in all wavelength channels are correlated.  Because we lack a full covariance matrix, we consider this error analysis a first estimate and will adjust the scale of the errors ellipses by a scalar factor later in the analysis as we fit the binary orbit. The results from this analysis can be found in Table~\ref{table:astrometry}. Note that 
because of different (u,v) coverages and seeing conditions, the errors vary strongly between the different nights.  Visibilities and closure phases from MIRC for UT 2012 Jun 15 and visibilities from PAVO for UT 2017 Jun 14 along with the best fit models are shown in Figure \ref{visplot}. 

The stellar angular diameters and flux ratio between components were poorly constrained on individual nights. To improve our estimate, we used the final orbit (derived in \S\ref{sec:rvorbit}) to allow a global fit for the diameters and flux ratio under the assumption they do not vary (although this is not strictly true because of the $\delta$ Scuti pulsations). From the orbit we fixed the orbital geometry and then fitted the angular diameters and flux ratio with the full dataset, using bootstrap sampling to estimate our errors.  Table \ref{table:stellarproperties} contains the results of this work: UD1 (brighter star) 0.49$\pm$0.03 mas, UD2 (fainter star) 0.49$\pm$0.03 mas, flux ratio 1.04$\pm$0.03. These errors include uncertainties on the wavelength scale ($\pm$0.25\%) and on the calibrator diameters.

To improve our diameter estimates, we also collected single-baseline observations of $\delta$~Del with the visible-light PAVO combiner.  The individual nights did not have sufficient (u,v) coverage to simultaneously constrain relative positions as well as the stellar properties.  Following a procedure similar to MIRC, we used the precise orbit predictions from our model to fix the orbital geometry for the 5 nights of PAVO observations.  We then did a global least-squares fit (and bootstrap) with the following free parameters: UD diameter of component 1, UD diameter of component 2, $R$-band flux ratio of component 1 over component 2.  The best-fit reduced $\chi^2$ was 2.5, higher than normal, which may be due to uncertainty in the wavelength scale of PAVO ($\pm$0.6\%) being unaccounted for.  Table\,\ref{table:stellarproperties} also contains the PAVO results: UD1 (brighter star) 0.460$\pm$0.014 mas, UD2 (fainter star) 0.510$\pm$0.014 mas, Flux ratio 1.10$\pm$0.05.  These errors include uncertainties on the wavelength scale ($\pm$0.6\%) and on the calibrator diameters (5\%). 

Lastly, we need to determine our final estimate of the effective temperatures for the two components of $\delta$~Del.  To do this we used Kurucz/Castelli models \footnote{Specifically, we used the tables found at: \texttt{https://www.oact.inaf.it/castelli/castelli/grids/gridp00k2odfnew/fp00k2tab.html} and
 \texttt{https://www.oact.inaf.it/castelli/castelli/grids/gridm05k2odfnew/fm05k2tab.html}.} \citep{castelli2004} to fit for the limb-darkening corrected $R$ and $H$ band diameters determined from interferometry, the interferometrically determined component flux ratios, and literature photometry $R = 4.17 \pm 0.05$ \citep{morel1978}, $H = 3.70 \pm 0.24$ \citep{2mass}. We found an acceptable fit with the following stellar parameters: Component 1: LD diameter 0.500$\pm$0.014 mas, Temperature 7440K$\pm$210K; Component 2: LD diameter 0.507$\pm$0.014 mas, Temperature 7110K$\pm$180K. These parameters along with physical radii, luminosity, and component $R$/$H$ magnitudes can be found in Table~\ref{table:stellarproperties}.  We will use these properties to create a HR diagram in \S\ref{sec:hr}.

\begin{table}[H]
\centering
\caption{$\delta$ Del Astrometry Data}
\label{table:astrometry}
\begin{tabular}{lcccccc}
\hline
\colhead{UT Date} & \colhead{MJD} & \colhead{sep (mas)}   & \colhead{P.A. ($^\circ$)} & \colhead{error major axis (mas)} & \colhead{error minor axis (mas)} & \colhead{error ellipse P.A. ($^\circ$)} \\
\hline
2011 Jul 15 & 55757.331 & 7.166 & 337.31 & 0.004 & 0.002 & 302 \\
2011 Jul 17 & 55759.323 & 6.448 & 345.80 & 0.003 & 0.001 & 319 \\
2012 Jun 10 & 56088.492 & 4.274 & 15.80 & 0.049 & 0.009 & 341 \\
2012 Jun 12 & 56090.483 & 2.961 & 46.79 & 0.008 & 0.002 & 64 \\
2012 Jun 15 & 56093.450 & 2.120 & 170.36 & 0.004 & 0.003 & 287 \\
2012 Jun 16 & 56094.503 & 2.750 & 206.10 & 0.033 & 0.005 & 40 \\
2012 Jun 20 & 56098.446 & 5.363 & 256.94 & 0.012 & 0.011 & 56 \\
2012 Sep 19 & 56189.215 & 8.449 & 295.20 & 0.015 & 0.009 & 276 \\
2012 Sep 20 & 56190.219 & 8.570 & 297.96 & 0.005 & 0.004 & 37 \\
2013 Jul 13 & 56486.512 & 7.662 & 331.29 & 0.07 & 0.016 & 38 \\
2013 Jul 14 & 56487.351 & 7.430 & 334.10 & 0.005 & 0.003 & 337 \\
\hline
\end{tabular}
\end{table}

\begin{table} [H]
\begin{center}
\caption{$\delta$ Del Stellar Properties}
\label{table:stellarproperties}
\begin{tabular}{lccc}
\tableline
\tableline
 & Component 1 & -- & Component 2 \\
\tableline
$f_1/f_2$ $_{\text{$H$-band}}$ & -- & $1.04 \pm 0.03$ & -- \\
$f_1/f_2$ $_{\text{$R$-band}}$ & -- & $1.10 \pm 0.05$ & -- \\
$H$ (mag) & $4.43 \pm 0.24$ & & $4.47 \pm 0.24$ \\
$R$ (mag) & $4.87 \pm 0.03$ & & $4.98 \pm 0.03$ \\
$\theta_\mathrm{UD}$ $_{\text{$H$-band}}$ (mas) & $0.49 \pm 0.03$ & & $0.49 \pm 0.03$ \\
$\theta_\mathrm{UD}$ $_{\text{$R$-band}}$ (mas) & $0.460 \pm 0.014$ & & $0.510 \pm 0.014$ \\
$\theta_\mathrm{LDD}$ (mas) & $0.500 \pm 0.014$ & & $0.507 \pm 0.014$ \\
Radii ($R_\odot$) & $3.43 \pm 0.11$ & & $3.48 \pm 0.11$ \\
Temperature (K) & $7440 \pm 210$ & & $7110 \pm 180$ \\
Luminosity ($L_\odot$) & $32.4 \pm 4.2$ & & $28.8 \pm 3.4$ \\
\tableline
\end{tabular}
\end{center}
\end{table}

\subsection{Spectroscopy}
\label{sec:spectroscopy}
We acquired 97 useful spectroscopic observations of $\delta$~Del between
2012 June and 2016 June with the Tennessee State University 2~m Automatic 
Spectroscopic Telescope (AST) and a fiber-fed echelle spectrograph 
\citep{ew07} that is located at Fairborn Observatory in southeast Arizona.
The detector was a Fairhild 486 CCD that has a 4096 $\times$ 4096 array of 15 
micron pixels. The echelle spectrograms have 48 orders that cover a wavelength
range from 3800 to 8260~\AA. Our observations were made with a fiber that
produces a resolution of 0.24~\AA, and the spectrograms have typical 
signal-to-noise ratios of 70--130. \citet{fetal13} have provided additional
information about the facility.
 
\begin{figure}[H]
\centering
\includegraphics[width=\linewidth]{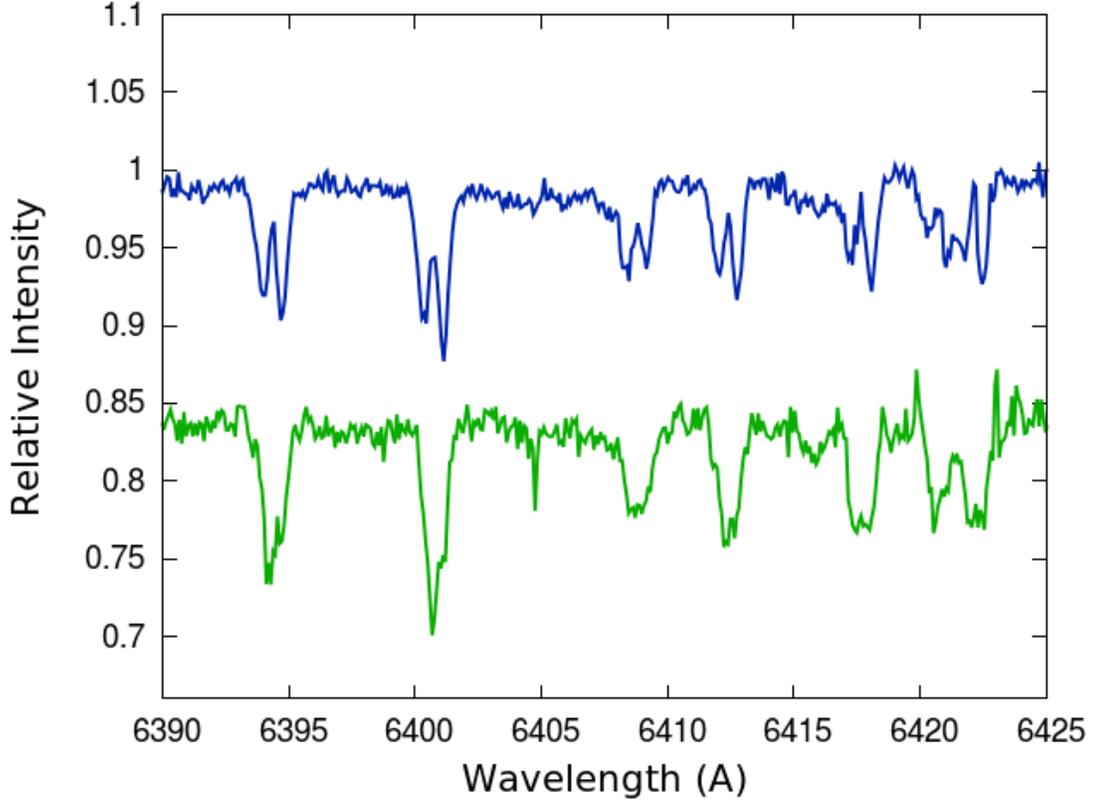}
\caption{Plotted are partial spectra from Fairborn Observatory for $\delta$ Del at maximum velocity separation and at lower velocity separation. At maximum velocity separation (top spectrum) the peaks from the two components are almost completely resolved. This gives us good constraints on the line width and depth, enabling reliable fits also at the epochs with smaller velocity separation, when the lines are blended (bottom spectrum).}
\label{spectrum}
\end{figure}

\citet{ftw09} gave a general explanation of the velocity measurement  
of the AST echelle spectrograms. For $\delta$~Del we used our solar-type
star line list that contains 168 lines in the wavelength range 4920--7100~\AA.
At our resolution the lines of the two components at maximum velocity separation
are almost completely resolved. At most other phases the features are 
very significantly blended as can be seen in Figure \ref{spectrum}. We used 
rotational broadening functions \citep{lf11, fg11} to fit simultaneously the 
line pairs. Because of pulsation, the shapes of the lines vary to some extent 
from spectrum to spectrum. Therefore, 
although we used the average width and depth values from our most widely 
separated line pairs as starting values for our velocity determinations, those two parameters were not fixed in our fits. To test for systematics affecting our radial velocity determinations for this blended system, we divided our line list into blue (4920-5501~\AA) and red (5506-7200~\AA) halves and remeasured velocities for 
6 spectra near maximum velocity separation and 10 spectra near the lower velocity separation. After comparing radial velocities determined from the red half, the blue half, and the full wavelength range, we see no striking systematics in our results. 

Our unpublished velocity measurements of several IAU radial velocity standards
from spectra obtained with our 2~m AST have an average velocity difference of 
$-$0.6~km~s$^{-1}$ when compared to the results of \citet{s10}. Thus, to each 
of our measured velocities we have added 0.6~km~s$^{-1}$. The 97 Fairborn radial 
velocities used for orbit fitting are listed in Table \ref{table:rv}. In addition to these velocities, we measured two single-lined spectra from the 2 m AST to determine velocities very close to the phase of the center-of-mass velocity. At MJD 56197.2428 we obtain a single-lined radial velocity of 10.4 km~s$^{-1}$, and at MJD 57090.5220 we obtain a velocity of 8.8 km~s$^{-1}$. We did not include these two points in the fitting routine because the precision of these measurements is lacking due to $\delta$ Scuti pulsations and different rotational velocities of the components. However, the positions of these single-lined velocities appear to support the system velocity and mass ratio obtained in our best fit orbit described in Section \ref{sec:rvorbit}. 

From our fits to the lines in our Fairborn Observatory spectra that
are at phases near maximum velocity separation, we have determined 
$v$~sin~$i$ values of 17 $\pm$ 1 km~s$^{-1}$ for the more massive 
primary star and 12 $\pm$ 1 km~s$^{-1}$ for the less massive 
secondary. For the same subset of spectra that we used to determine the 
$v$~sin~$i$ values of the components, we measured the average line
equivalent widths of the two stars. That ratio, which 
for stars of similar temperature corresponds to the luminosity
ratio of the components, was highly variable, likely because of the 
rather significant $\delta$~Scuti pulsations that also affect the
line profiles. With the ratio of the more massive primary to the 
less massive secondary ranging from 1.2 to 0.9, the average ratio 
is 1.03 $\pm$ 0.02 for a central wavelength of 6000~\AA. Thus we assume 
that the more massive star is also the brighter component henceforth. 

At the Lick Observatory 87 spectra of $\delta$ Del were obtained with the 120-inch telescope at a dispersion of 5.3 ~\AA~mm$^{-1}$ (Duncan 1973).  Ten lines in the wavelength range 3900-4300~\AA\ were used to determine radial velocities for both components. Velocity measurements were made with a Grant measuring engine and reduced 
with a standard computer program. These radial velocities, which only cover 
phases very close to maximum velocity separation, are presented in Table \ref{table:rvduncan} and have not been published until now. We use the radial 
velocities of both components from the 87 observations acquired at Lick Observatory, 
as well as the 97 new observations obtained at Fairborn Observatory when 
carrying out our orbital fitting routines. 

\begin{longtable}{lccc}
\caption{$\delta$ Del Radial Velocities from Fairborn Observatory} \label{table:rv}\\
\hline
MJD & $v_1$ [$\pm 1.3$] (km/s) & $v_2$ [$\pm 0.7$] (km/s) & $v_1$ ($\delta$ Scuti subtracted) \\
\hline
\endhead
56100.194 & 20.2 & -0.6 & 19.7 \\
56101.23 & 18.9 & 1.2 & 18.0 \\
56101.305 & 18.2 & 0.7 & 19.0 \\
56106.306 & 16.6 & 3.3 & 15.8 \\
56107.345 & 16.0 & 4.0 & 15.1 \\
56126.284 & -0.3 & 18.5 & 1.6 \\
56168.311 & -2.6 & 20.6 & -1.1 \\
56169.306 & 1.0 & 24.0 & -1.3 \\
56172.306 & -5.9 & 28.0 & -7.4 \\
56173.306 & -4.9 & 25.7 & -2.4 \\
56261.174 & 19.3 & -2.0 & 17.3 \\
56262.168 & 18.2 & -1.4 & 18.4 \\
56266.142 & 16.7 & 1.5 & 18.7 \\
56267.141 & 17.7 & 3.3 & 15.4 \\
56415.456 & -4.9 & 26.9 & -6.9 \\
56547.301 & 19.0 & -0.3 & 18.3 \\
56547.343 & 20.4 & -1.5 & 18.3 \\
56547.366 & 19.1 & -0.2 & 18.7 \\
56549.149 & 14.9 & 0.8 & 17.0 \\
56574.09 & 0.9 & 22.8 & 2.3 \\
56575.101 & 1.3 & 21.7 & -0.5 \\
56576.106 & -6.5 & 26.2 & -4.0 \\
56577.089 & -4.7 & 27.4 & -5.1 \\
56577.104 & -3.3 & 27.1 & -5.0 \\
56577.204 & -7.7 & 26.3 & -5.2 \\
56578.088 & -8.1 & 28.6 & -9.4 \\
56578.204 & -7.1 & 28.6 & -8.9 \\
56579.087 & -10.2 & 27.6 & -7.8 \\
56583.084 & 21.5 & -2.0 & 19.2 \\
56584.094 & 16.6 & -1.7 & 19.0 \\
56585.082 & 18.0 & -1.2 & 17.9 \\
56586.112 & 17.6 & -1.6 & 18.9 \\
56623.097 & 20.0 & 0.2 & 19.6 \\
56624.17 & 22.0 & -1.7 & 19.8 \\
56736.481 & -6.2 & 18.1 & -4.6 \\
56737.501 & -4.4 & 19.6 & -6.0 \\
56740.469 & -8.1 & 26.5 & -8.8 \\
56741.46 & -5.4 & 25.8 & -7.1 \\
56742.492 & 2.0 & 18.3 & 2.8 \\
56744.471 & 20.2 & 1.2 & 21.4 \\
56745.448 & 17.4 & -1.1 & 19.7 \\
56747.455 & 19.2 & 0.0 & 20.8 \\
56776.348 & -1.4 & 18.6 & 0.3 \\
56777.369 & 0.9 & 22.4 & -0.5 \\
56778.457 & 1.3 & 24.2 & -0.7 \\
56779.367 & -2.0 & 25.1 & -3.7 \\
56780.328 & -1.5 & 28.4 & -3.8 \\
56781.337 & -11.9 & 27.5 & -9.5 \\
56782.336 & -4.1 & 25.9 & -5.0 \\
56785.329 & 20.1 & -0.9 & 18.1 \\
56786.309 & 22.4 & -2.0 & 21.2 \\
56787.337 & 21.6 & -1.8 & 22.2 \\
56788.318 & 22.0 & -1.6 & 19.7 \\
56789.3 & 19.7 & -1.6 & 19.6 \\
56822.212 & -7.3 & 29.1 & -8.7 \\
56822.255 & -8.7 & 28.5 & -6.5 \\
56823.212 & -6.6 & 24.1 & -4.2 \\
56826.255 & 20.5 & -2.3 & 18.4 \\
56826.288 & 20.7 & -1.7 & 19.3 \\
56827.255 & 18.5 & -1.9 & 19.5 \\
56827.288 & 18.5 & -1.6 & 20.9 \\
56828.294 & 21.8 & -1.2 & 19.7 \\
56829.326 & 15.8 & -1.9 & 18.2 \\
56830.323 & 21.4 & -0.6 & 19.8 \\
56831.287 & 20.0 & -1.4 & 17.7 \\
56899.285 & -0.6 & 22.9 & 0.1 \\
56944.086 & -9.6 & 30.2 & -7.3 \\
56945.133 & 0.0 & 24.4 & -1.6 \\
56954.19 & 19.5 & 0.3 & 17.7 \\
57103.516 & -4.9 & 23.1 & -4.6 \\
57115.506 & 18.5 & -0.8 & 18.8 \\
57143.489 & -2.2 & 21.9 & -2.7 \\
57184.291 & -5.2 & 23.1 & -2.9 \\
57185.354 & -3.7 & 24.9 & -4.1 \\
57186.354 & -9.5 & 27.8 & -7.5 \\
57187.355 & -6.4 & 28.6 & -8.7 \\
57188.355 & -7.1 & 25.8 & -5.5 \\
57192.375 & 22.7 & -0.8 & 20.4 \\
57347.068 & -2.8 & 25.7 & -0.8 \\
57348.075 & -2.9 & 25.6 & -5.2 \\
57349.082 & -9.6 & 27.0 & -7.2 \\
57350.069 & -9.3 & 28.4 & -9.2 \\
57351.07 & 0.2 & 24.8 & -1.4 \\
57356.067 & 20.4 & -2.9 & 18.1 \\
57511.352 & -9.5 & 26.2 & -7.6 \\
57512.316 & -10.2 & 29.0 & -7.8 \\
57513.334 & 0.1 & 24.6 & -2.2 \\
57516.307 & 20.6 & -0.1 & 18.3 \\
57517.306 & 18.9 & -0.7 & 20.6 \\
57518.304 & 20.2 & -1.8 & 20.6 \\
57519.34 & 18.2 & -1.4 & 19.5 \\
57520.329 & 19.1 & -0.6 & 20.6 \\
57551.217 & -8.3 & 26.5 & -6.8 \\
57552.219 & -5.1 & 26.8 & -7.4 \\
57553.221 & -10.5 & 27.5 & -8.3 \\
57557.219 & 23.3 & -0.6 & 21.2 \\
57558.243 & 15.8 & -2.2 & 18.2 \\
\hline
\end{longtable}

\begin{longtable}{lcccc}
\caption{$\delta$ del RVs obtained from Lick Observatory} \label{table:rvduncan}\\
\hline
MJD & $v_1$ [$\pm 2.6$] (km s$^{-1}$) & $v_2$ [$\pm 1.3$] (km s$^{-1}$) & $v_1$ ($\delta$ Scuti subtracted) & $v_2$ ($\delta$ Scuti subtracted)\\
\hline
\endhead
38306.242 & -9.5 & 29.9 & -9.1 & 28.5 \\
39238.541 & -7.6 & 25.6 & -5.5 & 26.6 \\
39239.409 & -5.8 & 28.3 & -9.1 & 27.0 \\
39239.442 & -6.8 & 26.4 & -8.2 & 26.9 \\
39239.452 & -8.8 & 27.1 & -9.1 & 28.0 \\
39239.467 & -10.8 & 27.4 & -9.6 & 28.4 \\
39239.474 & -11.1 & 27.7 & -9.3 & 28.5 \\
39239.51 & -7.7 & 29.4 & -6.4 & 28.3 \\
39239.516 & -8.0 & 28.4 & -7.3 & 27.1 \\
39239.522 & -7.1 & 28.4 & -7.0 & 27.0 \\
39279.447 & -8.3 & 28.9 & -7.2 & 28.6 \\
39280.303 & -5.2 & 28.4 & -6.8 & 28.0 \\
39280.317 & -2.9 & 30.3 & -5.7 & 30.7 \\
39280.359 & -2.3 & 30.2 & -4.2 & 30.7 \\
39280.362 & -6.1 & 30.2 & -7.7 & 30.5 \\
39280.373 & -9.0 & 28.9 & -9.5 & 28.6 \\
39280.382 & -8.6 & 29.6 & -8.0 & 28.8 \\
39280.39 & -8.2 & 29.7 & -6.9 & 28.5 \\
39280.398 & -8.0 & 30.2 & -6.1 & 28.8 \\
39280.408 & -8.4 & 32.5 & -6.2 & 31.0 \\
39280.415 & -8.6 & 33.6 & -6.4 & 32.3 \\
39280.423 & -9.8 & 32.1 & -7.8 & 31.0 \\
39280.43 & -10.5 & 30.9 & -9.0 & 30.2 \\
39280.436 & -10.8 & 29.9 & -9.9 & 29.6 \\
39280.442 & -11.2 & 29.7 & -10.9 & 29.7 \\
39280.45 & -9.9 & 29.0 & -10.4 & 29.4 \\
39280.456 & -9.0 & 27.9 & -10.1 & 28.6 \\
39280.462 & -8.3 & 27.8 & -10.1 & 28.7 \\
39281.356 & -8.8 & 28.3 & -6.6 & 27.6 \\
39362.232 & -9.8 & 28.8 & -8.7 & 29.7 \\
39362.247 & -11.6 & 28.2 & -9.5 & 28.5 \\
39362.259 & -10.4 & 29.0 & -8.2 & 28.6 \\
39362.27 & -9.0 & 28.7 & -7.2 & 27.7 \\
39362.282 & -0.7 & 29.5 & 0.1 & 28.1 \\
39362.298 & -4.3 & 28.4 & -5.2 & 27.0 \\
39362.312 & -4.3 & 27.3 & -6.6 & 26.5 \\
39362.324 & -4.5 & 25.9 & -7.6 & 25.7 \\
39362.335 & -5.6 & 28.6 & -8.9 & 29.0 \\
39362.346 & -5.4 & 27.1 & -8.4 & 28.0 \\
39362.359 & -5.2 & 27.2 & -7.2 & 28.2 \\
39362.373 & -7.0 & 25.6 & -7.6 & 26.2 \\
39362.386 & -9.2 & 27.8 & -8.4 & 27.7 \\
39362.41 & -12.5 & 27.7 & -10.3 & 26.4 \\
39362.422 & -11.8 & 28.3 & -9.8 & 26.8 \\
39362.435 & -9.1 & 27.2 & -7.9 & 25.9 \\
39401.111 & -5.9 & 25.4 & -5.3 & 26.4 \\
39401.132 & -3.5 & 26.9 & -1.4 & 27.6 \\
39401.152 & -6.2 & 28.5 & -4.3 & 28.1 \\
39401.256 & -7.2 & 27.0 & -7.9 & 27.9 \\
39401.269 & -8.1 & 27.0 & -7.4 & 27.5 \\
39401.282 & -6.4 & 26.9 & -4.6 & 26.6 \\
39401.34 & -6.4 & 26.8 & -7.4 & 26.1 \\
39402.098 & -10.9 & 29.5 & -9.3 & 28.2 \\
39402.165 & -3.2 & 27.3 & -6.4 & 28.2 \\
39402.175 & -5.3 & 27.9 & -8.1 & 28.9 \\
39402.222 & -9.5 & 29.2 & -7.7 & 28.1 \\
39402.231 & -9.7 & 29.6 & -7.5 & 28.2 \\
39402.239 & -7.9 & 30.0 & -5.7 & 28.5 \\
39402.319 & -3.3 & 27.8 & -6.6 & 28.6 \\
39402.333 & -3.3 & 29.2 & -6.0 & 29.4 \\
39726.278 & -2.1 & 24.2 & -3.2 & 25.2 \\
39726.284 & -3.0 & 24.3 & -3.5 & 25.2 \\
39726.289 & -7.1 & 25.3 & -7.0 & 26.1 \\
39726.297 & -7.4 & 27.0 & -6.5 & 27.4 \\
39726.305 & -9.3 & 27.5 & -7.7 & 27.5 \\
39726.31 & -9.9 & 27.8 & -8.0 & 27.5 \\
39726.316 & -10.9 & 27.6 & -8.8 & 27.0 \\
39726.322 & -9.8 & 28.0 & -7.6 & 27.0 \\
39726.323 & -9.9 & 28.9 & -7.7 & 27.9 \\
39726.333 & -9.5 & 28.2 & -7.5 & 26.8 \\
39726.338 & -8.6 & 29.3 & -6.9 & 27.8 \\
39726.344 & -6.0 & 28.9 & -4.7 & 27.4 \\
39726.351 & -5.2 & 29.7 & -4.5 & 28.3 \\
39726.365 & -3.4 & 29.8 & -4.3 & 29.0 \\
39726.384 & -5.5 & 26.1 & -8.2 & 26.4 \\
39726.396 & -3.3 & 26.3 & -6.6 & 27.1 \\
39726.406 & -3.6 & 25.3 & -6.9 & 26.3 \\
39726.421 & -2.0 & 26.2 & -4.5 & 27.0 \\
39726.438 & -5.2 & 26.8 & -6.0 & 26.8 \\
39727.345 & -7.4 & 29.4 & -10.7 & 30.2 \\
39727.426 & -11.8 & 28.6 & -9.6 & 27.9 \\
39727.438 & -9.3 & 28.2 & -7.8 & 28.3 \\
39727.451 & -6.1 & 28.5 & -5.7 & 29.2 \\
39728.252 & -9.1 & 24.5 & -10.6 & 25.4 \\
40781.310 & -10.8 & 30.5 & -12.9 & 29.3 \\
40782.270 & -10.2 & 29.7 & -10.3 & 28.7 \\
\hline
\end{longtable}
\clearpage

\section{Orbit Fitting Routine}
\label{sec:orbit}
\subsection{Astrometry Model}
The Campbell elements ($\omega$,$\Omega$,$e$,$i$,$a$,$T$,$P$) describe the motion of one star of a binary system relative to the other. Those symbols have their usual meanings where $\omega$ is the longitude of the periastron, $\Omega$ is the position angle of the ascending node, $e$ is the eccentricity, $i$ is the orbital inclination, $a$ is angular separation, $T$ is a time of periastron passage, and $P$ is orbital period. Good overviews for the use of least-squares fitting to determine the best fit orbital elements are given by \citet{Wright2009} and \citet{Lucy2014}. The errors in our positions for $\delta$ Del are ellipses, and thus, to determine the best fit orbital elements with a least-squares routine, we must project the residuals into the major and minor ellipse axes when defining $\chi^2$. We define $\chi^2$ in the major and minor axes as

\begin{equation}
\begin{aligned}
\chi_{major}^2=\frac{[(x_{data}-x_{model})\sin{\sigma_{pa}}+(y_{data}-y_{model})\cos{\sigma_{pa}}]^2}{\sigma_{major}^2}\\
\chi_{minor}^2=\frac{[-(x_{data}-x_{model})\cos{\sigma_{pa}}+(y_{data}-y_{model})\sin{\sigma_{pa}}]^2}{\sigma_{minor}^2},
\end{aligned}
\end{equation}
where $\sigma_{pa}$, $\sigma_{major}$, and $\sigma_{minor}$ are the error ellipse position angle, error in major axis, and error in minor axis, respectively. The final positions predicted by our model are given by $x_{model}$ and $y_{model}$, while $x_{data}$ and $y_{data}$ are the positions measured by MIRC. The total $\chi^2$ for the astrometry data is then just the sum of $\chi^2_{major}$ and $\chi^2_{minor}$. The reduced $\chi^2$ for our best fit suggests that astrometry error values are overestimated. We reduce the error values by a factor of $\sim 3.5$ to bring the reduced $\chi^2$ to 1. This ensures that one dataset is not unevenly weighting the fitting when combining astrometry and radial velocity data. 

\subsection{Radial Velocities Model}
The orbital elements for the double-lined spectroscopic binary are $\omega$, $e$, $K_1$, $K_2$, $\gamma$, $T$, and $P$. The elements $\omega$, $e$, $T$, and $P$ are the same as presented in the astrometry model, $K_1$ and $K_2$ are the velocity semi-amplitudes of each component, and $\gamma$ is the systemic velocity. These elements are used to compute a model value for velocity at each time of observation. Once again the total $\chi^2$ for radial velocity data is just the sum of the individual components, $\chi^2_{primary}$ and $\chi^2_{secondary}$. The Fairborn radial velocities cover a 
much more extensive portion of the full orbit, and each velocity is the average of a much greater number of lines, so we assign these velocities twice the weight of those obtained at Lick Observatory. Since the reduced $\chi^2$ for the radial velocity best fit is $>1$, we increase the RV error values by a factor of $\sim 1.3$ to bring reduced $\chi^2$ to 1 for fitting the combined RV and astrometry data. 

\subsection{Fitting Methods}
We use a Markov chain Monte Carlo (MCMC) fitting routine to determine the best fit parameters for our binary model. An MCMC fit can efficiently sample a large region of parameter space to ensure that a global minimum has been reached, unlike the least-squares method which can become stuck at local minima solutions. Parameter distributions from the MCMC sampling also provide more accurate error values than those obtained via least-squares fitting. We carry out an MCMC fit using the Python package \textit{emcee} developed by \citet{Foreman2013}. 

Assuming independent Gaussian errors for our data, the log-likelihood function is just given as
\begin{equation}
ln(\mathcal{L})=-\frac{1}{2}\chi^2,
\end{equation}
where $\chi^2$ is formulated as explained in the previous subsections. As a starting point for our MCMC walkers we use the Python package \textit{lmfit} for non-linear least squares fitting \citep{Newville2014}. To sample a large amount of parameter space, we randomly perturb each parameter about its best fit value from least-squares as a starting point for each walker. For each fit we run 2*N$_{params}$ walkers until convergence is reached. The Gelman-Rubin diagnostic \citep{gelman1992} is used to test whether or not a chain has converged. This diagnostic compares the variance of a parameter in one chain with the variance between chains and is given by 
\begin{equation}
R=\sqrt{\frac{Var(\Theta)}{W}},
\end{equation}
where $\Theta$ is some parameter and $W$ is the variance within a single chain. As a chain converges, this ratio approaches 1. In the results presented, all of our chains have been run until R$<$1.001. Uniform priors are used for each parameter, with search range restrictions given in Table \ref{table:mcmcparams}. 

\begin{table}[H]
\centering
\caption{MCMC Parameter Search Range}
\label{table:mcmcparams}
\begin{tabular}{lcc}
\hline
\colhead{Parameter} & \colhead{Min Value ($>$)} & \colhead{Max Value ($<$)} \\
\hline
P (days) & 40 & 41 \\
T (MJD) & 56823 & 56825 \\
e & 0 & 1 \\
$\omega$ ($\deg$) & 0 & 360 \\
$\Omega$ ($\deg$) & 0 & 360 \\
i ($\deg$) & 0 & 180 \\
a (mas) & 5.0 & 6.0 \\
K$_1$ (km/s) & 0 & 30 \\
K$_2$ (km/s) & 0 & 30 \\
$\gamma$ (km/s) & 0 & 30 \\
\hline
\end{tabular}
\end{table}

The orbital elements can be determined from separate fits to astrometry and RV data or from combining the datasets to fit all ten orbital parameters at once. In the next section we present fitting results for all three cases (astrometry alone, RV alone, and combined fit). The results of our fitting routines for $\delta$ Del are presented in the next section.

\section{Orbital Fitting Results}
\label{sec:fitting_results}
\subsection{Astrometry Alone}
\label{sec:astrometryalone}
\begin{figure}[H]
\centering
\includegraphics[width=4in]{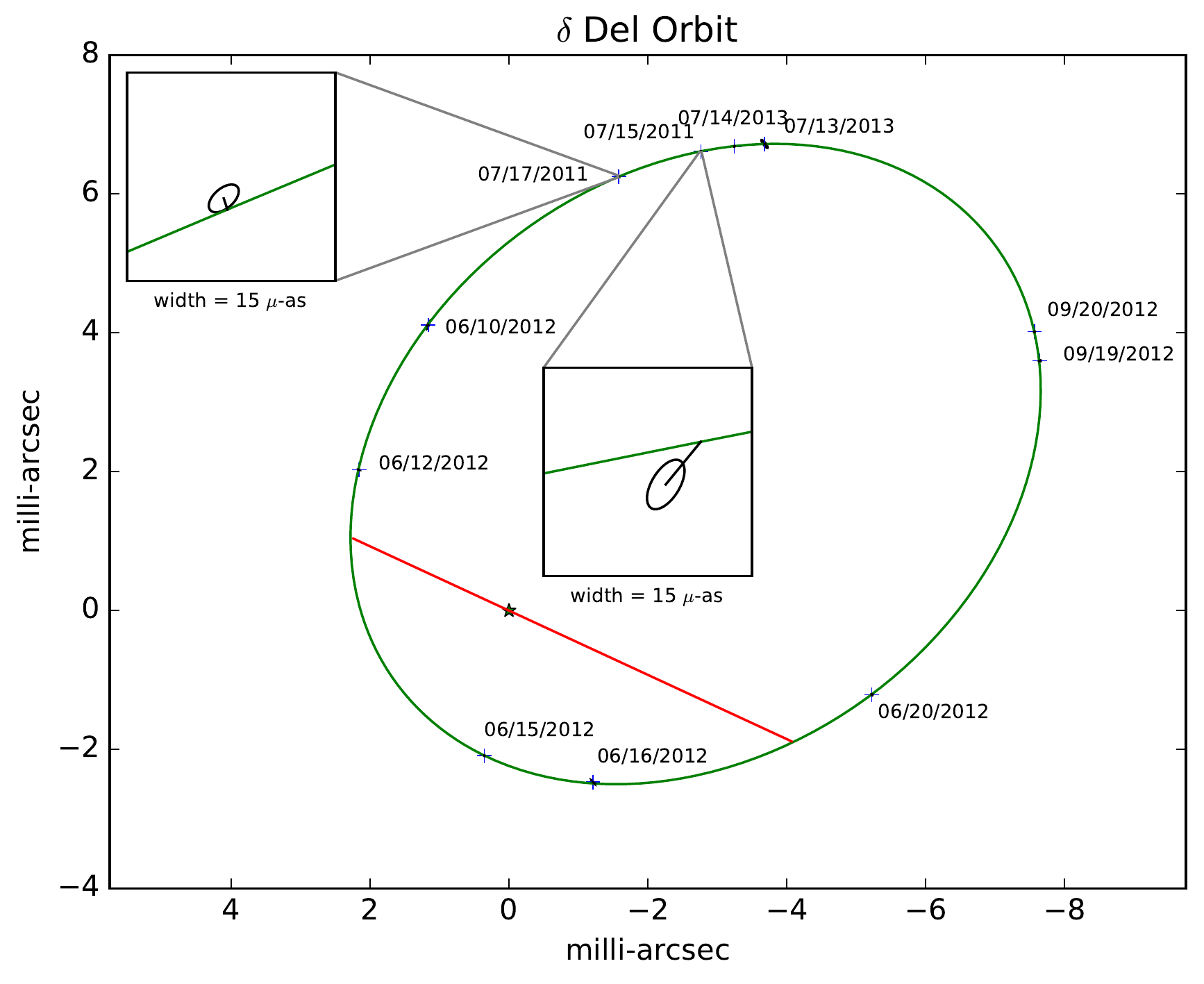}
\caption{The position of the primary component relative to its companion at the origin. The green ellipse is the best fit orbit for $\delta$ Del determined by fitting to astrometry data. The error ellipses on each observation are too small to be seen at this scale, so the insets display representative errors. The red line is the line of nodes, about which the orbit is inclined. Orbital motion is counterclockwise, with the portion of the orbit above the line of nodes inclined towards the observer and the portion below the line inclined away.}
\label{orbitplot}
\end{figure}

\begin{figure}[H]
\centering
\includegraphics[width=\linewidth]{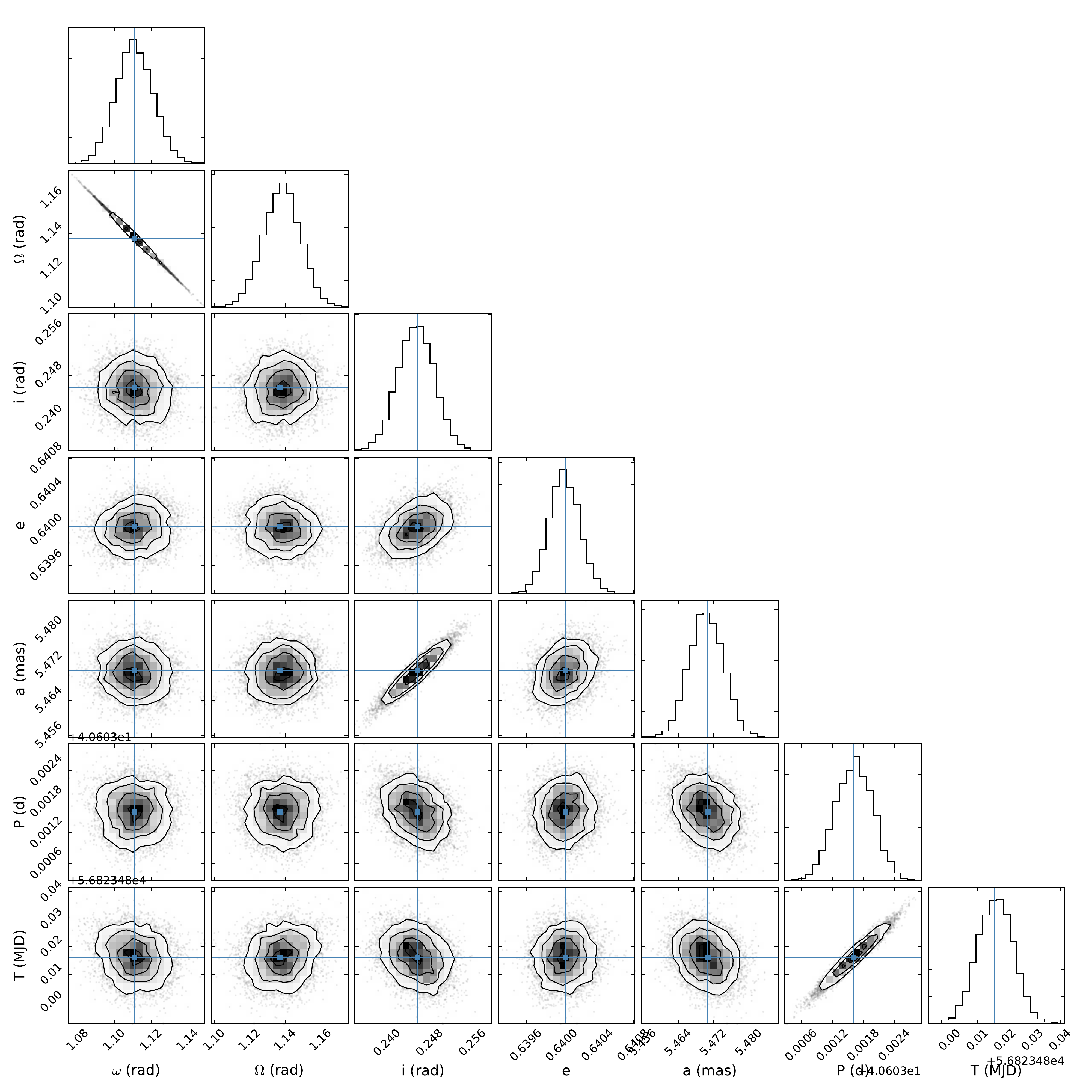}
\caption{A corner plot of parameter distributions from the MCMC routine for astrometry data. Histograms display the number of times a given value was chosen as the best value for that element, and 2D plots show correlations between parameters. The crosshairs denote the best fit from least-squares fitting.}
\label{astrometrymcmc}
\end{figure}

Using our described fitting routine we first determine the best fit orbital elements from astrometry data alone. The best fit orbit along with our measured positions is shown in Figure \ref{orbitplot}. Also plotted is the line of nodes, about which the binary orbit is inclined. Data points near the nodes are crucial for constraining the angular semi-major axis, while points away from the nodes help constrain the inclination. The best fit parameters and their errors from MCMC fitting are displayed in Table \ref{orbitelements}. Figure \ref{astrometrymcmc} shows parameter posterior distributions. Correlations between $T$ and $P$, $\omega$ and $\Omega$, and $a$ and $i$ are expected from a visual orbit. Our quoted error bar on each parameter is the standard deviation of the posterior distribution from the MCMC routine. Along with the MCMC error, there is a systematic error of $\pm 0.25\%$ on the angular semi-major axis due to MIRC absolute wavelength calibration \citep{monnier2012}.

\subsection{Radial Velocity Alone}
\label{sec:rvorbit}
\begin{figure}[H]
\centering
\includegraphics[width=5in]{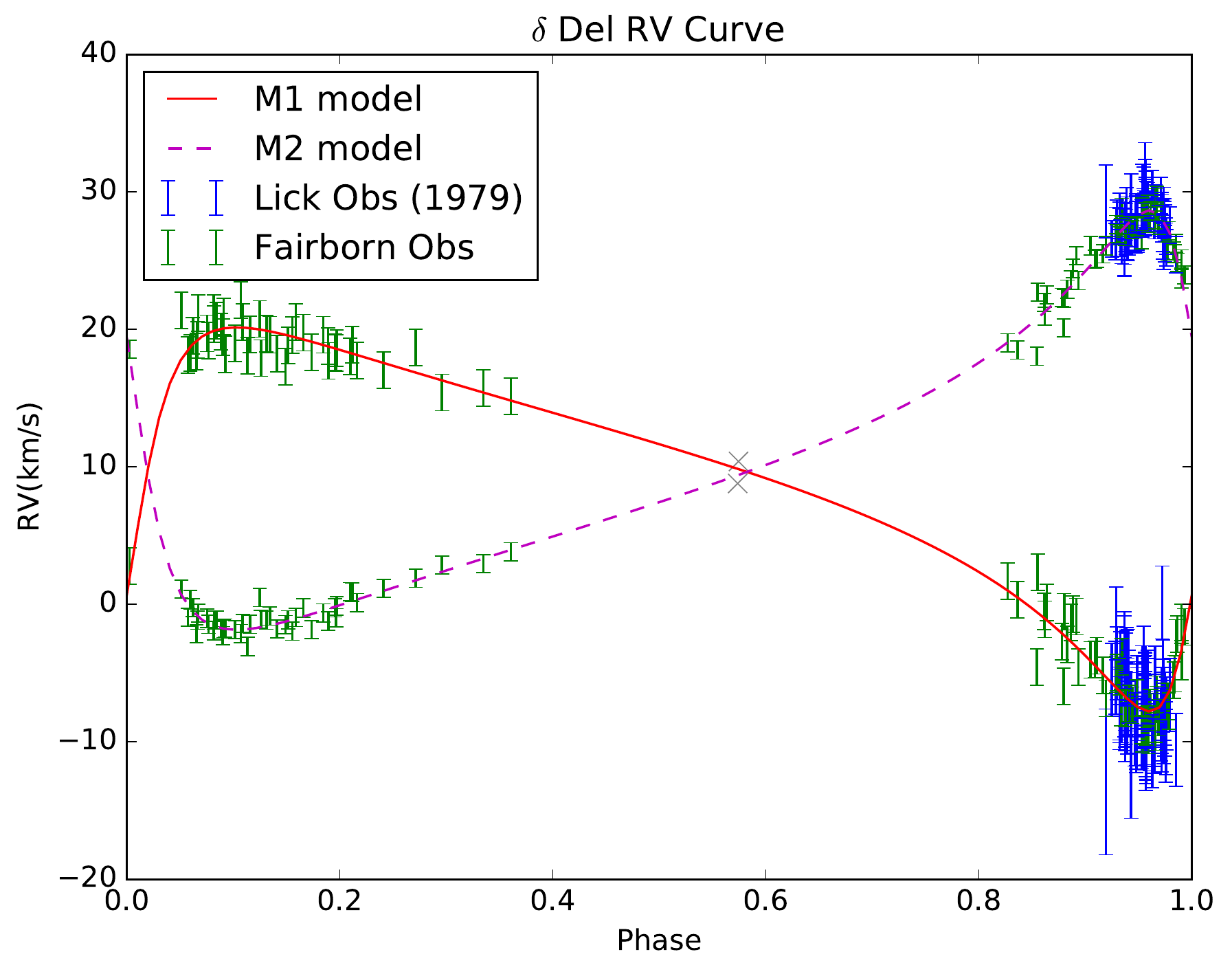}
\caption{Double-lined radial velocities along with the best fit model. The RV data combines 97 data points from Fairborn Observartory with \cite{Duncan1979}'s 87 unpublished RVs from Lick Observatory. The two gray marks just before phase 0.6 are velocities obtained from single-lined spectra. These points are not included in the best fit, but the measurements do help support our system velocity and mass ratio values.}
\label{rvbest}
\end{figure}

\begin{figure}[H]
\centering
\begin{minipage}[b]{.45\textwidth}
\centering
\includegraphics[width=\linewidth]{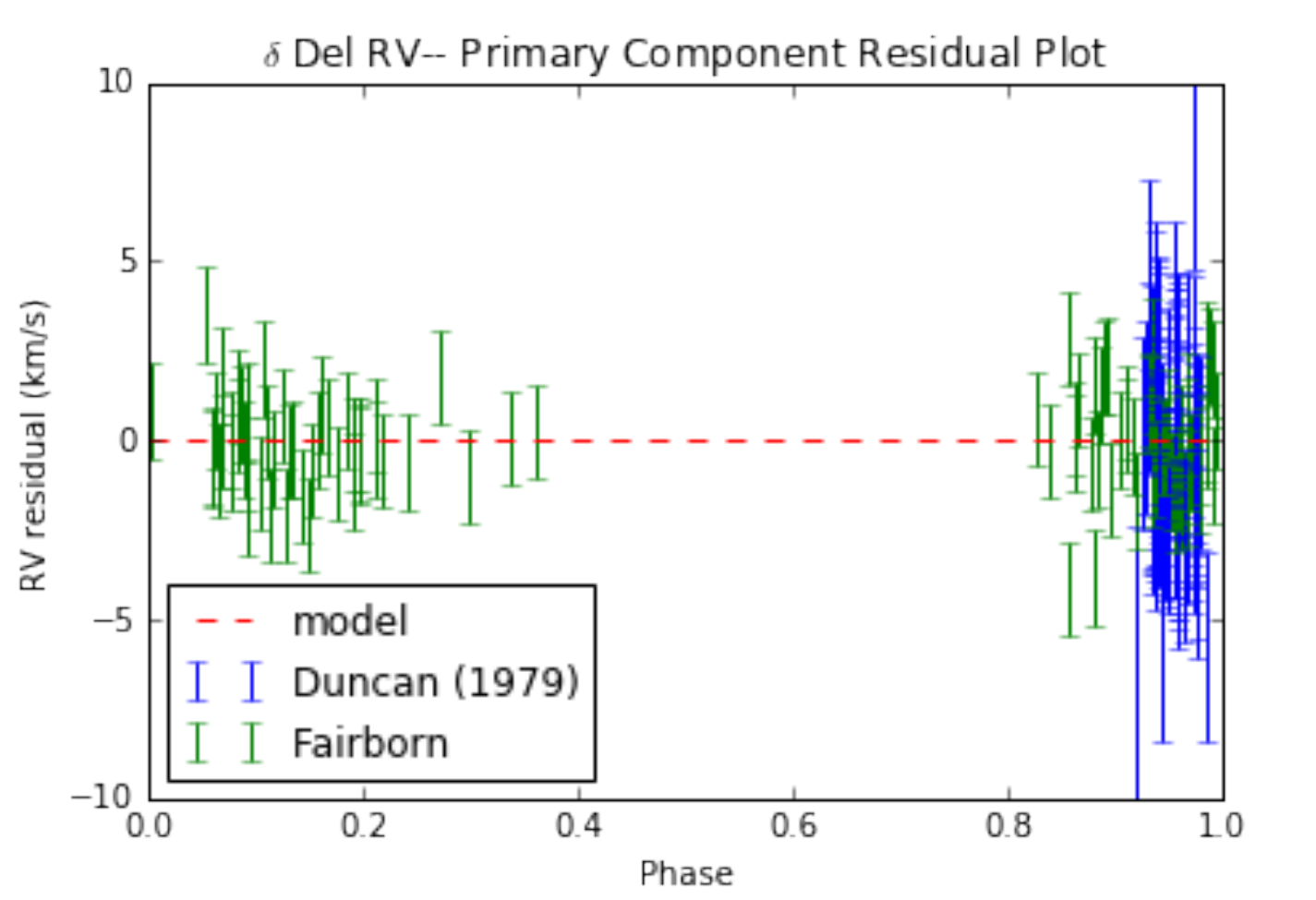}
\caption{Residual plot of the primary component of $\delta$ Del from the best fit RV orbit.}
\label{primaryresid}
\end{minipage}
\hfill
\begin{minipage}[b]{.45\textwidth}
\centering
\includegraphics[width=\linewidth]{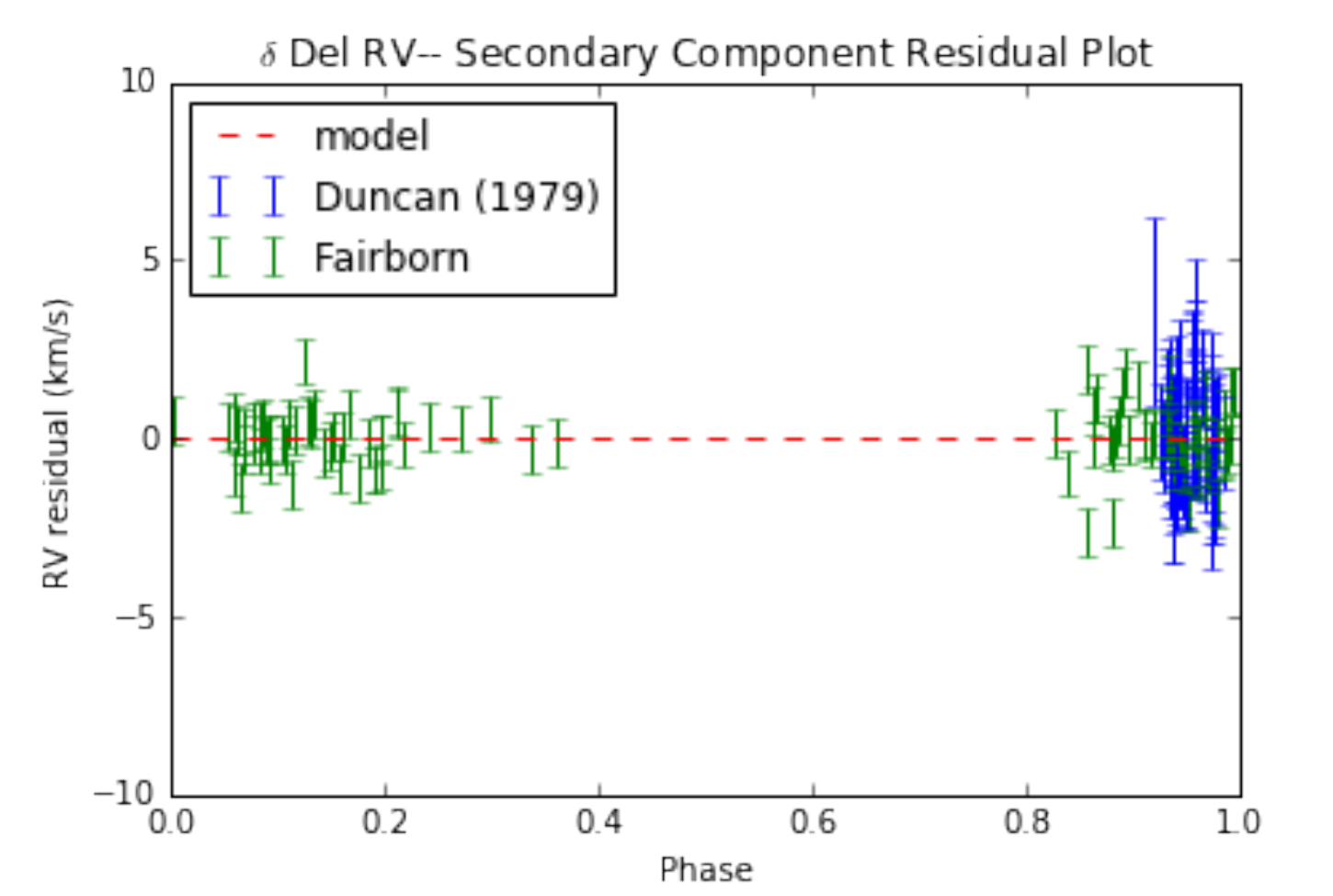}
\caption{Residual plot of the secondary component of $\delta$ Del from the best fit RV orbit.}
\label{secondaryresid}
\end{minipage}
\end{figure}

\begin{figure}[H]
\centering
\includegraphics[width=\linewidth]{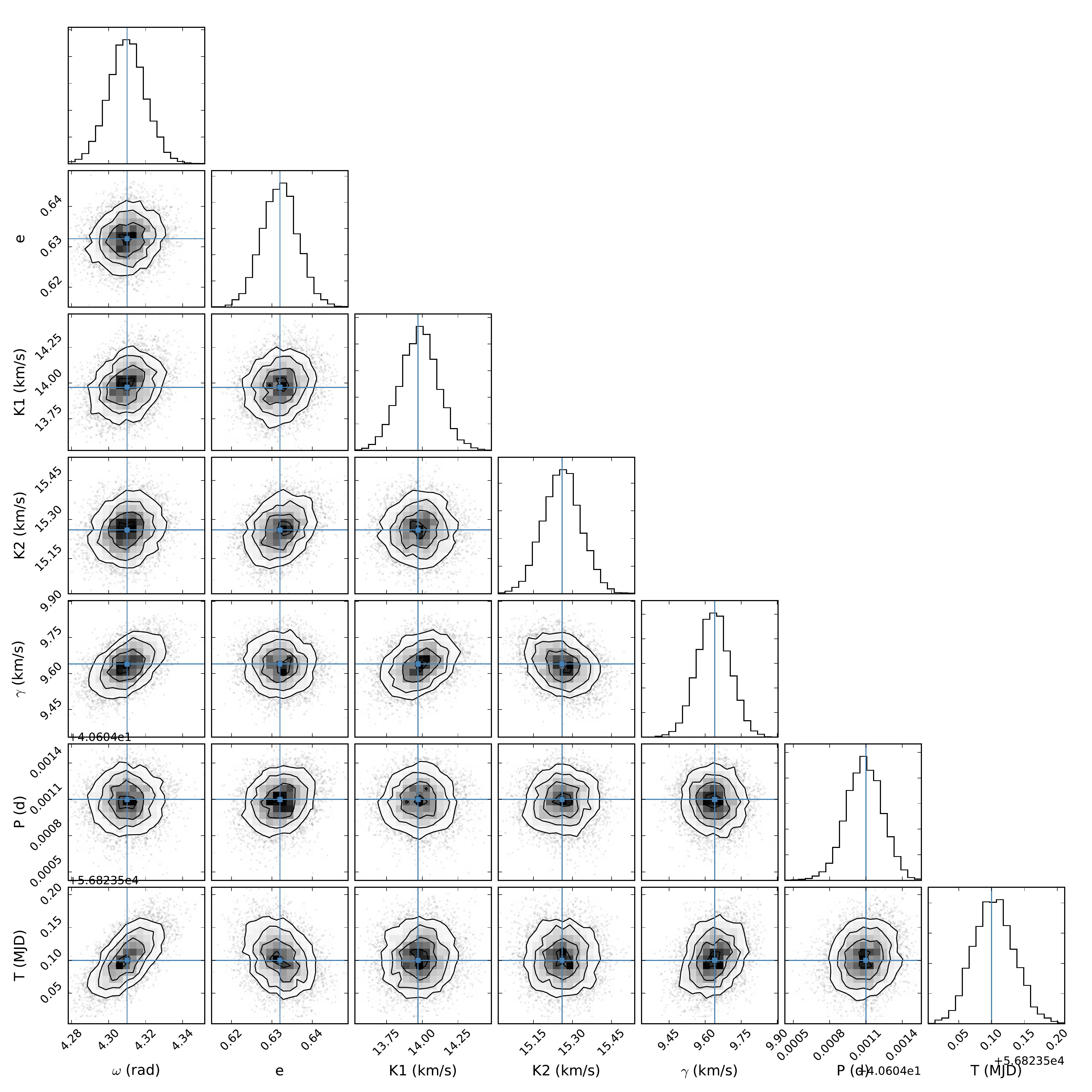}
\caption{A corner plot of parameter distributions from the MCMC routine for RV data. Histograms display the number of times a given value was chosen as the best value for that element, and 2D plots show correlations between parameters. The crosshairs denote the best fit from least-squares fitting.}
\label{rvmcmc}
\end{figure}

Due to short period variations in their radial velocity curves, both components of $\delta$ Del have been previously classified as $\delta$~Scuti variables with periods of $0.158 \pm 0.006$ days for the primary (more massive) component and $0.134 \pm 0.015$ days for the secondary \citep{Duncan1979}. Though modeling these pulsations does not change the final orbital solution, we do detect $\delta$ Scuti variations in portions of our data. We detect significant period signals for the primary component in the Fairborn and Lick Observatory data, as well as for the secondary component in the Lick Observatory data. We describe our first-order corrections for these pulsations in Appendix \ref{appendix}, and we list the resulting corrected RVs in Tables \ref{table:rv} and \ref{table:rvduncan}. Once the $\delta$ Scuti pulsations are subtracted out of the RV data, we determine the best fit orbital elements to the RV data alone using our MCMC routine. Figure \ref{rvbest} shows our best fit orbit with residual plots shown in figures \ref{primaryresid} and \ref{secondaryresid}. These include the 97 double-line RV points from Fairborn Observatory as well as the 87 data points from Lick Observatory. We also plot the two velocities measured from single-lined spectra near phase 0.6. These velocities are not included in the fit, due to the low precision of these points. However, the two velocities appear to support our best fit values of system velocity and hence mass ratio. Figure \ref{rvmcmc} displays parameter posterior distributions. Table \ref{orbitelements} shows the best fit orbital elements from fitting to RV data alone, along with MCMC error values. \citet{Duncan1979} reported values of $40.580 \pm 0.003$ and $0.7 \pm 0.1$ for the period and eccentricity in their preliminary orbit analysis. Our best value of 40.6051 $\pm$ 0.0002 days for the orbital period differs slightly from theirs, while our eccentricity of 0.632 $\pm$ 0.004 
is within their quoted uncertainty. 

\subsection{Combined Fit with Physical Orbital Parameters}
\begin{figure}[H]
\centering
\includegraphics[width=\linewidth]{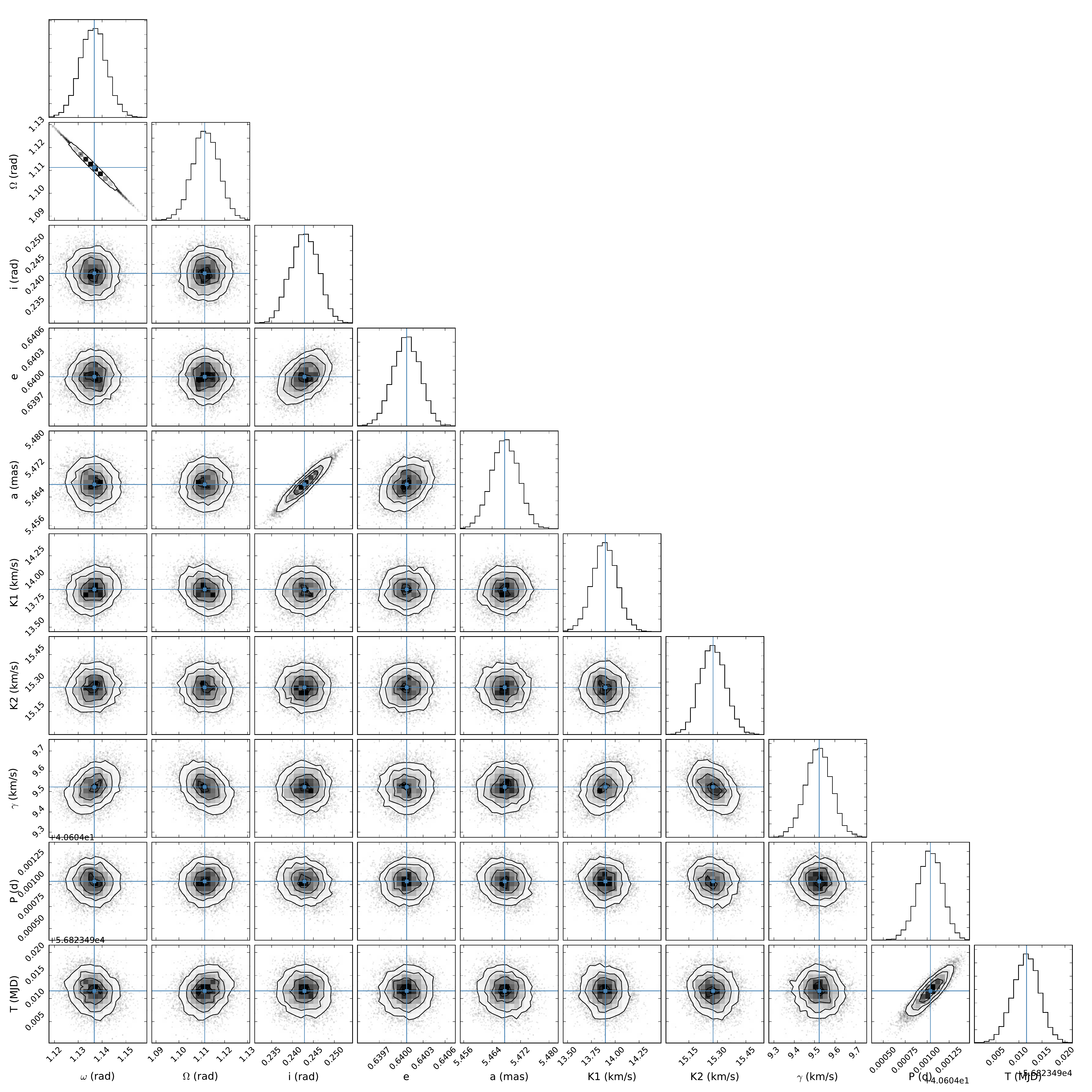}
\caption{A corner plot of parameter distributions from the MCMC routine for combined astrometry and RV data. Histograms display the number of times a given value was chosen as the best value for that element, and 2D plots show correlations between parameters. The crosshairs denote the best fit from least-squares fitting.}
\label{combinedmcmc}
\end{figure}

Since orbital elements $\omega$, $e$, $T$, and $P$ are constrained by both astrometry and RV data it is advantageous to combine the datasets for a single fit. When combining datasets we assign a weight to each set to bring both reduced $\chi^2_\mathrm{ast}$ and $\chi^2_\mathrm{RV}$ to 1 when fitting separately. The total $\chi^2$ to be minimized is 

\begin{equation}
\chi_\mathrm{combined}^2=w_\mathrm{ast}*\chi_\mathrm{ast}^2+w_\mathrm{RV}*\chi_\mathrm{RV}^2,
\end{equation}
where $w_{ast}$ and $w_{RV}$ are the weights assigned to the astrometry and radial velocity datasets. Table \ref{orbitelements} shows the best fit values for all ten orbital parameters determined from fitting to the combined set of data. Figure \ref{combinedmcmc} shows the parameter distribution from our MCMC fitting routine. Note that there is a $\pm 0.25\%$ wavelength calibration systematic error on the angular semi-major axis as mentioned in section \ref{sec:astrometryalone}. This systematic error affects the distance value determined from the orbit. 

Combining astrometry and RV data leads to a measurement of physical orbital elements of parallax, linear semi-major axis, and masses of each component (see \citet{torres2010} for relevant equations). These values and their errors are shown in Table \ref{physicalelements}. Our results agree with the original parallax measurement by Hipparcos of $16.03 \pm 0.68$ mas \citep{Perryman1997}. However, the revised Hipparcos reduction for $\delta$ Del reports a parallax of $14.61 \pm 0.2$ mas \citep{vanLeeuwen2007}, which is not consistent with our measurement. Our new 
parallax measurements decreases the Hipparcos distance of $\delta$ Del from $68.45 \pm 0.94$ pc to our new value of $63.61 \pm 0.89$ ($\pm 0.16$ systematic error) pc. Since Hipparcos did not identify this source as a binary there could be systematic errors in the parallax determination, since photocenter motion due to binarity could effect the parallax fit. However, since the magnitudes are nearly equal in R band one would not expect a large photocenter shift. We point out that a discrepancy from the revised Hipparcos reduction has been reported before in the close binary system $\psi$ Persei \citep{mourard2015}. 

Note that we present the results of fits carried out from velocities with the $\delta$ Scuti pulsations subtracted. However, we also carried out a combined fit using the measured RVs without
the $\delta$ Scuti RV variations subtracted. None of the orbital elements, mass ratio, or masses changed outside of the error bars quoted in the best fit solution with the $\delta$ Scuti variations subtracted.

\begin{table}[H]
\centering
\caption{Best fit orbital elements from astrometry and RV data}
\label{orbitelements}
\begin{tabular}{llll}
\hline
\colhead{} & \colhead{Astrometry Alone} & \colhead{RV Alone} & \colhead{Astrometry+RV} \\
\hline
$P$ (d) & $40.60510 \pm 0.00015$ & $40.60514 \pm 0.00016$ & $40.60505 \pm 0.00014$ \\
$T$ (MJD) & $56823.604 \pm 0.030$ & $56823.6180 \pm 0.032$ & $56823.5019 \pm 0.0028$ \\
$e$ & $0.6319 \pm 0.0043$ & $0.6334 \pm 0.0046$ & $0.64008 \pm 0.00018$ \\
$\omega (^\circ)$ & $67.17 \pm 0.58$ & $66.94 \pm 0.61$ & $65.07 \pm 0.32$ \\
$\Omega (^\circ)$ & $65.17 \pm 0.58$ & -- & $63.73 \pm 0.33$ \\
$i (^\circ)$ & $14.08  \pm 0.19$ & -- & $13.92\pm 0.18$ \\
$a$ (mas) & $5.4707 \pm 0.0039$\tablenotemark{1} & -- & $5.4676 \pm 0.0037$\tablenotemark{1} \\
$K_1$ (km/s) & -- & $13.98 \pm 0.14$ & $13.88 \pm 0.14$ \\
$K_2$ (km/s) & -- & $15.26 \pm 0.07$ & $15.27 \pm 0.07$ \\
$\gamma$ (km/s) & -- & $9.61 \pm 0.07$ & $9.48 \pm 0.07$ \\
\hline
\end{tabular}
\tablenotetext{1}{$\pm 0.014$ (systematic)}
\end{table}

\begin{table}[H]
\centering
\caption{Best fit physical elements from combining RV and astrometry data}
\label{physicalelements}
\begin{tabular}{ll}
\hline
\colhead{Physical Element} & \colhead{Best Value} \\
\hline
parallax, $\pi$ (mas) & $15.72 \pm 0.22$ ($\pm$0.04)\tablenotemark{1} \\
distance, $d$ (pc) & $63.61 \pm 0.89$ ($\pm$0.16)\tablenotemark{1} \\
semi-major axis, $a$ (AU) & $0.348 \pm 0.005$ \\
$M_1/M_2$ & $1.100 \pm 0.012$ \\
$M_1$ ($M_{\odot}$) & $1.78 \pm 0.07$ \\
$M_2$ ($M_{\odot}$) & $1.62 \pm 0.07$ \\
\hline
\end{tabular}
\tablenotetext{1}{systematic error in parentheses}
\end{table}

\section{Stellar Evolution for $\delta$ Del}
\label{sec:hr}
\subsection{Rotational Velocities and Orbital Evolution}
From our fits to the lines in our Fairborn Observatory spectra that are at phases near maximum velocity separation, we have determined $v$~sin~$i$ values of 17 $\pm$ 1 km~s$^{-1}$ for the more massive primary star and 12 $\pm$ 1 km~s$^{-1}$ for the less massive secondary. If the rotational and orbital axes are parallel, as is usually assumed, then we can use our orbital inclination value of 13.9\arcdeg\ to determine the equatorial rotational velocities of the components. With that inclination the projected velocities increase to 71 and 50 km~s$^{-1}$, respectively. 

Over time the orbits of close binaries tend toward circularization and rotational synchronization with the orbital period occurs for the components \citep[e.g.,][]{zahn1977,tassoul1987,tassoul1992,matthews1992}. In the case of an eccentric orbit, \citet{hut1981} has shown that the rotational angular velocity of a star will tend to synchronize with that of the orbital motion at periastron, a condition called pseudosynchronous rotation. With the periastron separation used as the semimajor axis, a period of 8.78 days results. Our computed radii from \S\ref{sec:interferometry} then produce pseudosynchronous velocities of 19.6 and 20.2 km~s$^{-1}$. Both values are much smaller than our equatorial rotational velocities. Given the youth of the system, its moderate orbital period, and that neither star has a significant outer convective envelope, it is not surprising that the rotational velocities of the components have not decreased to their pseudosynchronous values. 

\citet{gray1989}, \citet{gray2001}, and others have classified the composite spectrum of $\delta$~Del as a peculiar early F star, and \citet{reimers1976} found that its two components have identical peculiar chemical compositions. Such findings are consistent with the computed equatorial rotational velocities of the two stars, which are both less than 120 km~s$^{-1}$, the value below which A and early F stars generally have peculiar metal abundances \citep{abt1995}.

\subsection{Position on HR Diagram}
\begin{figure}[H]
\centering
\begin{minipage}[b]{.45\textwidth}
\centering
\includegraphics[width=\linewidth]{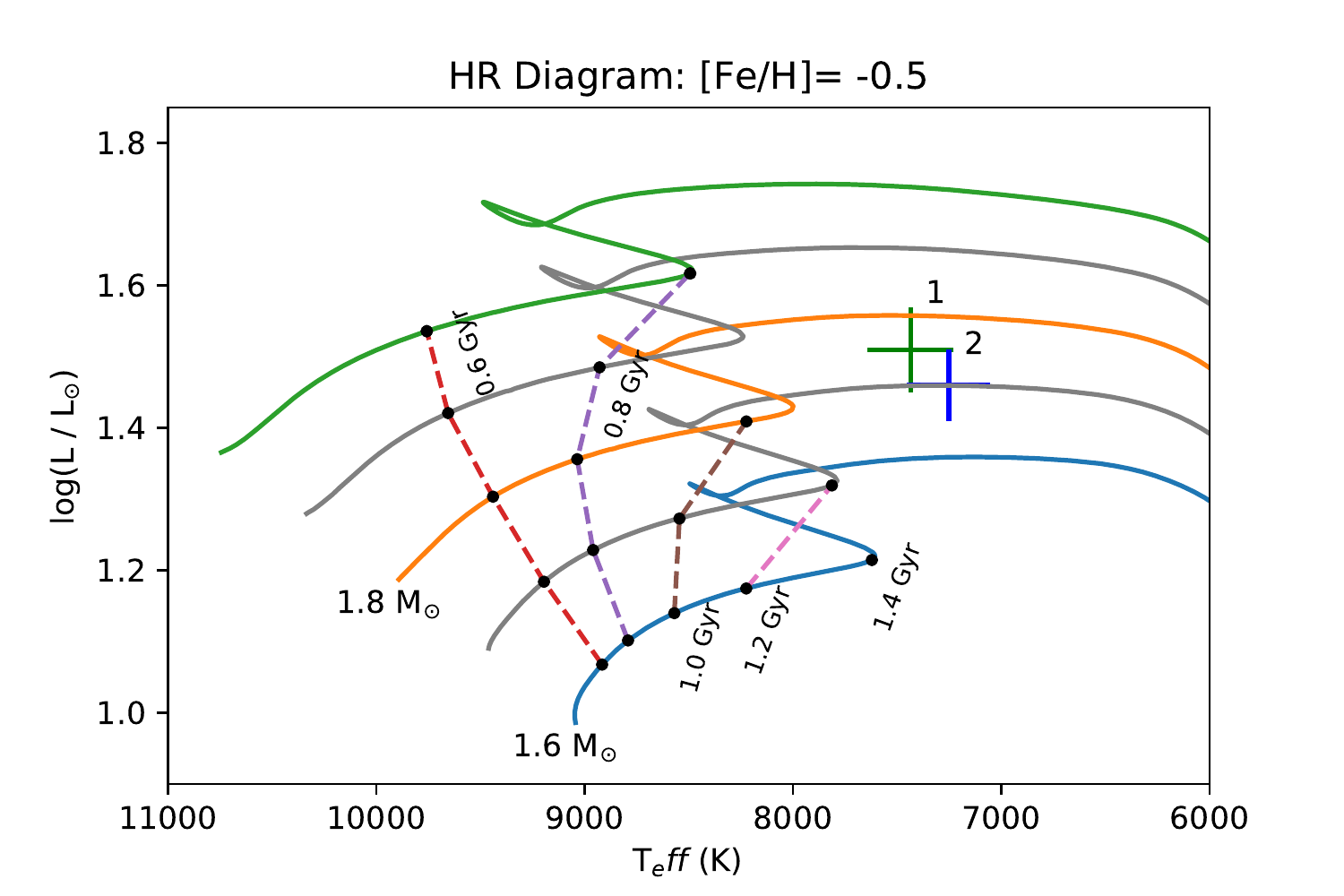}
\caption{Position of both components of $\delta$ Del on an HR diagram for the low metallicity case. MIST models are used to compute the plotted stellar tracks (solid lines) and isochrones (dashed lines). The luminosity and temperature values are consistent with our determined masses if component 2 is more evolved than component 1.}
\label{hr_lowmet}
\end{minipage}
\hfill
\begin{minipage}[b]{.45\textwidth}
\centering
\includegraphics[width=\linewidth]{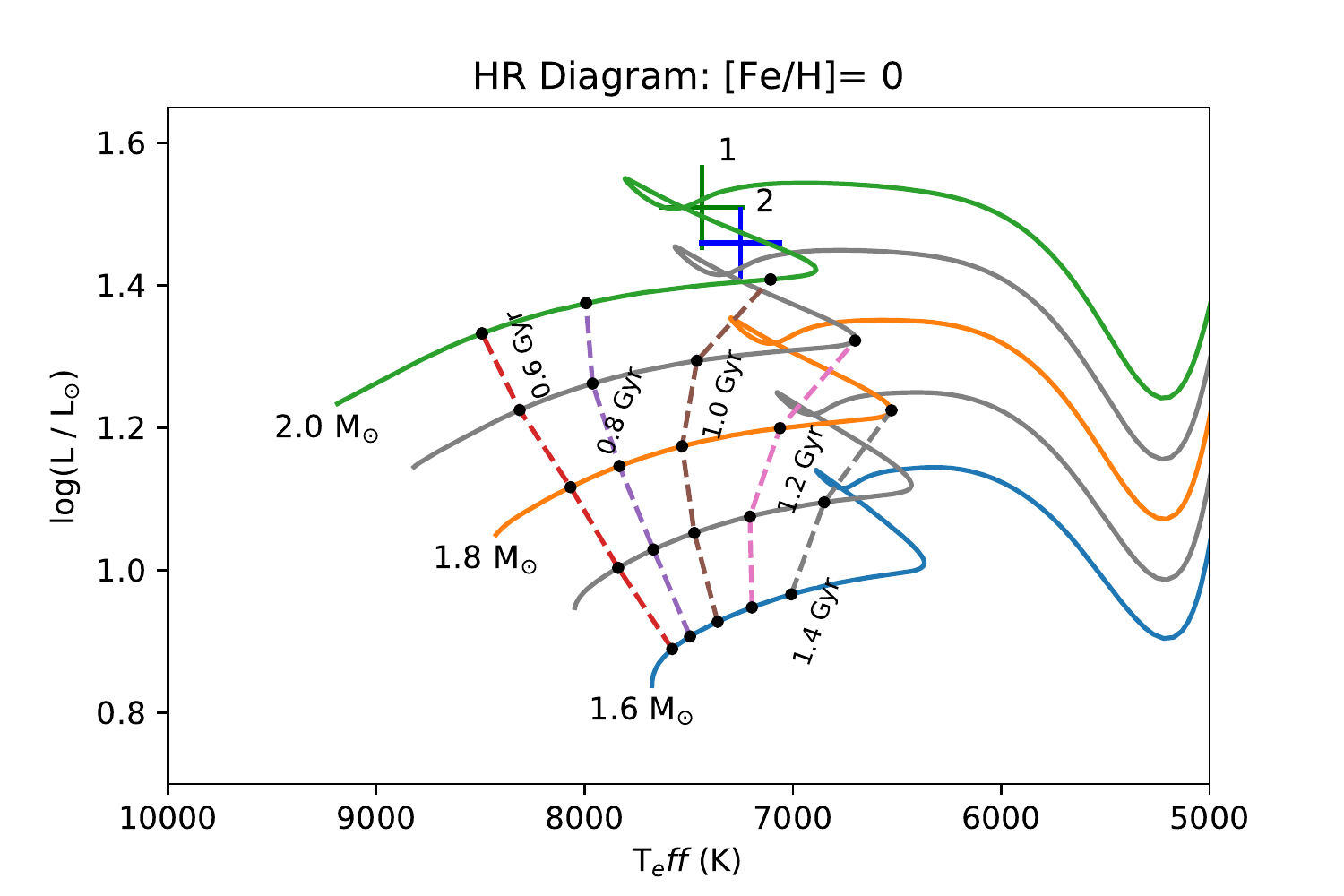}
\caption{Position of both components of $\delta$ Del on an HR diagram for the solar metallicity case. MIST models are used to compute the plotted stellar tracks (solid lines) and isochrones (dashed lines). The masses predicted from the HR diagram in this case are not consistent with those determined from our orbit.}
\label{hr_highmet}
\end{minipage}
\end{figure}

With our measured radii and flux ratios from MIRC and PAVO data, we are able to plot the position of both components of $\delta$ Del on an HR diagram. We use MESA Isochrones and Stellar Tracks (MIST) models to plot isochrones and tracks for different stellar masses \citep{mist1,mist2,mist3,mist4,mist5}. When compared with solar metallicity tracks, the track masses that match our luminosity and temperature determinations are not consistent with our best fit masses of $1.78$ $M_{\odot}$ and $1.62$ $M_{\odot}$ from our orbit. However, the metallicities for $\delta$ Del listed on SIMBAD suggest that this system may be metal poor. There is a spread in metallicity measurements from solar to metal poor values, depending largely on the adopted value of the effective temperature. \citet{reimers1976} measure [Fe/H]=$-$0.35, and \citet{cenarro2007} report [Fe/H]=$-$0.30. We find that a value of [Fe/H]=$-$0.5 gives solar tracks which are most consistent with our mass, luminosity, and temperature determinations. The position of each component of $\delta$ Del on an HR diagram, along with stellar tracks and isochrones, are shown for both low and solar metallicities in Figures \ref{hr_lowmet} and \ref{hr_highmet}. The mean H$\beta$ value, $b-y$, and $B-V$ colors listed in SIMBAD suggest a mean spectral class of about F0, which is what was found by \citet{morgan1972} and \citet{gray1989}. The best luminosity class estimates indicate that the average component for $\delta$ Del is evolved, consistent with our HR diagram results. Thus, the stars have evolved to late A or early F-type positions and were most likely originally late A-type stars. 

As can be seen from Figure \ref{hr_lowmet}, the individual masses determined from orbital fitting of radial velocity and astrometry data are only consistent with the measured radii and flux ratios if one stellar component is more evolved than the other. Note that although the error bars in Figure \ref{hr_lowmet} seem to overlap, the mass ratio above unity measured from the spectroscopic orbit makes overlap impossible. The position of the lower mass star on the HR diagram suggests an age $>1.2$ Gyr, while the age of the higher mass star is just over $1$ Gyr. This $\sim 200$ Myr age difference is puzzling, as one would expect two stars of a close binary system to be the same age. Since the components of $\delta$ Del are only separated at maximum RV separation, one possibility for this odd HR diagram placement is that there are systematic errors present in our radial velocity results which affect the mass ratio. The properties of the two stars derived from interferometry suggest that the mass ratio should be very close to unity, while our measured value from the spectroscopic orbit is $\sim 1.1$. It is not clear in which direction possible systematics would change the semi-amplitudes and, hence, mass ratio. The situation is further complicated by the pulsation of both components. Though we do not see any obvious systematics from our test described in \S\ref{sec:spectroscopy}, we nevertheless caution that systematic errors of the RV semi-amplitudes are a possible explanation for the odd positions of the components in the HR diagram. The two single-lined radial velocities that we measure when both components are at their center-of-mass velocity add further support that our value for system velocity is correct. This strengthens the claim of the mass ratio from the RV orbit, though we reiterate that these two velocity measurements are of low precision due to $\delta$ Scuti pulsations and different rotational velocities of the components that will not in general average out. 

Assuming that there are no systematic errors present in the mass ratio, we can think of four possible explanations for resolving the age difference problem in the HR diagram: 1) $\delta$ Scuti stars age differently than normal stars on the immediate post-main-sequence branch, 2) stellar evolution models are not accurate on the subgiant branch, 3) early interaction with a third component caused a difference in evolution rates, or 4) the age difference in the components of $\delta$ Del is a result of a merger event for the inner stars of an initially triple system. 

Theoretically, $\delta$ Scuti stars are expected to evolve as normal stars on the main-sequence and immediate post-main-sequence \citep[e.g][]{baglin1973,breger1979,breger1980}. However, as pointed out by \citet{peterson1996}, there is very little observational proof of this hypothesis. Recently \citet{niu2017} used photometric and spectroscopic data on the $\delta$ Scuti variable AE Ursae Majoris to provide such evidence that $\delta$ Scuti variables do in fact evolve as normal stars on the immediate post-main-sequence. However, one observation may not be sufficient for making this claim about all $\delta$ Scuti variables. A potential cause of abnormal aging among $\delta$ Scuti variables is the non-solar metal abundances present at the photosphere \citep[e.g.][]{guzik1998}. As pointed out by \citet{north1997} metallicity determination of $\delta$ Scuti variables may only be confined to the superficial layers of these stars and not reflect an internal metal distribution. Thus, mass determination via standard solar-scaled models may be invalid for these stars. \citet{tsvetkov1990} compared three different types of mass determinations for 89 $\delta$ Scuti variables. Although the mass determined from the evolutionary state on the HR diagram was consistent for most of their sample, for 9 of their stars the different methods of mass determination 
produced inconsistent results. The mass determination via the HR diagram differed 
by a factor of 2--5 between other methods. Hence, the HR diagram may not be reliable for mass determination for $\delta$ Scuti variables. \citet{north1997} also note that there is no one-to-one relation between mass and position on an HR diagram at the end of the core-hydrogen exhaustion phase. We find that $\delta$ Del lies right around this phase in stellar evolution, which may account for the discrepancies between mass prediction from the MIST stellar model and from the combined spectroscopic and visual orbit.

Close binary star evolution is in general a complex topic, where the closest systems often involve formation scenarios where the systems interact with a tertiary companions \citep{tokovinin2004}, and interaction with the circumstellar and circumbinary disks means that stars can be born with a variety of initial rotational velocities. Differential rotational velocities change interior mixing, and can cause  a difference in evolutionary rates. Additionally, interaction (such as accretion of He-rich material) with a now-ejected initially higher mass companion could also cause a difference in the evolutionary states between the two components. 

If the MIST models do in fact correctly describe these components, then the low-mass component must have an age of just over 1.2 Gyr while the high-mass component has an age of just over 1.0 Gyr. A possible way to account for this age difference is to assume that one of the stars is the result of a merger event. An inner binary of an initially triple system would have had to merge within $\sim 200$ Myr. The result of the merger would be a single star (the more massive component) which then evolved normally within the now binary system. The merger hypothesis has been proposed before to explain the existence of peculiar stars, and merger timescales of 100-500 Myr are theoretically possible \citep[e.g.][]{andrievsky1997,mink2014}. However, there are two major problems with one component of $\delta$ Del being the result of a merger event: 1) merger products are likely to have abnormal rotation rates, and 2) merger products are not likely to have a non-affected nearby main-sequence companion \citep{mink2014}. $\delta$ Del has both a relatively slow rotation rate and a very nearby companion. It is beyond the scope of this paper to determine whether or not it is truly possible for one component of this close binary system to be the result of an early merger event. Although it seems to be an unlikely scenario, if the stellar evolution models are correct for this binary then interaction with an early third companion is the only possibility we can think of to resolve the age discrepancy seen in the HR diagram.

\section{Toward Astrometric Detection of Exoplanets} 
\label{sec:exoplanets}

From the ground, long-baseline interferometry is a promising method for using differential astrometry to detect exoplanets. The astrometric detection method favors planets farther from the host star, unlike RV or transit surveys. Moreover, interferometric binary observations favor hot (A and B-type) binary stars which are difficult to probe via RV surveys because of weak and broad spectral lines. Thus, developing the capability to detect exoplanets with the MIRC instrument can probe a region that is not well explored by other detection methods. The recent PHASES project monitored binary stars with the Palomar Testbed Interferometer to obtain precise differential astrometric orbits and detected 6 candidate substellar objects orbiting single stars of a binary system \citep{Muterspaugh2010}. Unfortunately this project was halted due to the closure of the Palomar Testbed Interferometer in 2009. In this section, we demonstrate that the MIRC instrument at CHARA is capable of achieving the precision necessary for astrometric detection of exoplanets. The precision needed to detect the wobble of a star at $\delta$ Del's distance from a Jupiter mass planet within a few AU is on order of 10 $\mu$-arcseconds. With our $\delta$ Del orbit MIRC has achieved this precision in differential position of one star in a binary system with 10 minute observations. Thus, if there was a large planet around one component of $\delta$ Del it would be possible to detect as residuals on our astrometric orbit. Claiming a detection is not simple, as it involves adding 7 planet orbital parameters to the 7-parameter binary model. Adding free parameters to a model may lower the $\chi^2$ of a fit, but this does not necessarily make it a "better" model. A detection criterion often used for claiming radial velocity planet detections is the Bayesian Information Criteria (BIC) value \citep[e.g.][]{Feng2016,Motalebi2015,Sato2015}. The BIC is computed by
\begin{equation}
BIC=-2\ln{\mathcal{L}}+k\ln{n},
\end{equation}
where $k$ is the number of free parameters, $n$ is the number of data points, and $\mathcal{L}$ is the likelihood function. For our models, $-2\ln{\mathcal{L}}= \chi^2$. When comparing two models the one with a lower BIC value is selected as being a better fit to the data. 

We do not detect a planet around either component of $\delta$ Del, which is unsurprising since the binary separation is $\sim0.3$ AU. Still, we can use the precision of this orbit to test planet detection limits around $\delta$ Del and gain insight as to the types of planets we can detect when extending this precision to wider binary systems. To compute detection limits, we add simulated planet wobbles to our observations and fit the resulting data with a binary fit and a binary+planet fit. Note that we are testing for planets around individual stars of a binary system. while it is possible that a circumbinary planet exists around $\delta$ Del, our differential astrometric data is not sensitive to these types of orbits. We also emphasize that in this study we are only testing which planets show statistically significant detection signals with our measurement precision. Sophisticated fitting routines and many epochs of observations will be needed to recover the full orbit of real planets. Though fitting to 14 free parameters is a formidable challenge, in reality we will target systems where the 7 binary parameters are known quite well. Thus, only the 7 planet orbital elements will truly be free parameters. Future work of our group will include developing such fitting routines, building off of the work of recent studies that have tackled this challenge \citep[e.g.][]{Perryman2014,sozzetti2014,Ranalli2017}. 

The position of one star plotted relative to the companion is a sum of the position due to the binary orbit and the perturbation from the planet. Relative to a star at the origin, we can calculate the position vector $[x_s(t),y_s(t)]$ of one companion. We can also calculate the perturbation on a star due to an orbiting planet. Using the planet orbital elements, the position vector of the star from the planet is $[x_p(t),y_p(t)]$. The final astrometric position of a star with a binary companion and orbiting planet is then a sum of the two vectors
\begin{equation}
[x(t),y(t)]=[x_s(t),y_s(t)]+\frac{[x_p(t),y_p(t)]}{1+M_s/M_p},
\end{equation}
where $M_s$ is the mass of the star and $M_p$ the planet mass. The planet vector is shortened since we are only seeing the reflex motion of the star due to the presence of the planet. 

\begin{figure}[H]
\centering
\includegraphics[width=5in]{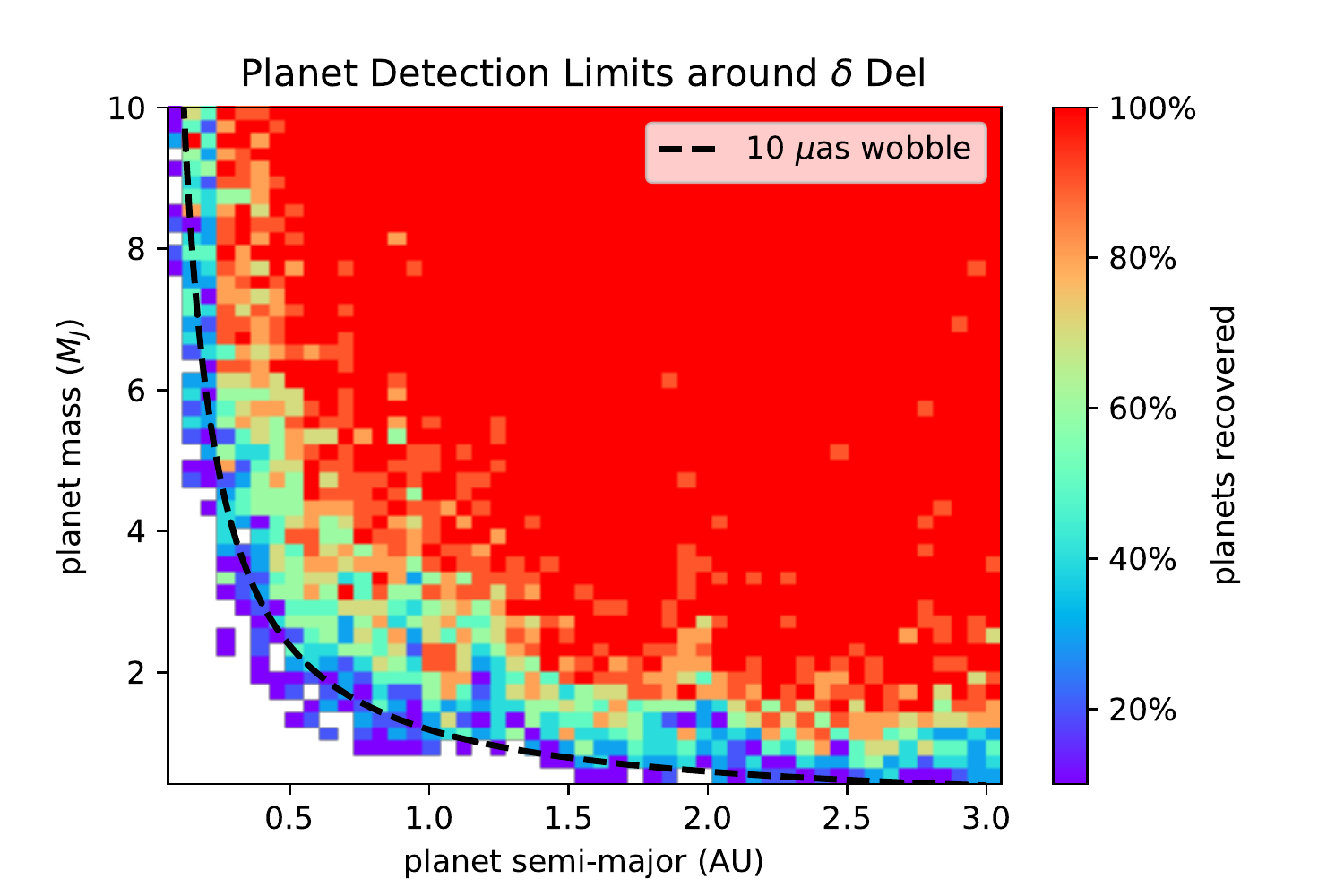}
\caption{Planet detection limits around $\delta$ Del are computed by simulating the wobble from a planet at each point on a mass, semi-major grid. A candidate detection is made based on the BIC criterion. If the best fit planet mass and semi-major axis are within $30$\% of the actual input values then we classify it as a true detection. For each mass, semi-major grid point we simulate 10 planets with random orbital elements and record the percentage of the time the planet is recovered. The dashed curve denotes the points on the grid where a planet would impart a 10 micro-arcsecond wobble on the star.}
\label{bic}
\end{figure}

To test planet detection limits around $\delta$ Del we simulate 10 planets with 0 eccentricity and random values for $\omega$, $\Omega$, $i$, and $T_0$ at each point on a grid with semi-major axes varying from $0.01-3$ AU and masses from $0.01-10$ $M_J$. We record the percentage of the planets we successfully recover at each grid point. The planet perturbation at the time of data collection is added to each real data point of our $\delta$ Del orbit. For each simulated planet we perform a binary fit (7 parameters) and a binary+planet fit (14 parameters) and compare the BIC values. We use the known binary and simulated-planet parameters as initial guesses for a least-squares fit to compute $\chi^2$ for the binary and binary+planet model. The model including a planet in the binary system is considered better if it has a lower BIC value and $\Delta$BIC$>$5 between the models \citep{liddle2007}. We consider true detections to be those in which the recovered planet mass and semi-major axis are within 30\% of the true input values of the simulated planet. Figure \ref{bic} displays our planet detection limits around a binary star with the observational precision of $\delta$ Del. Our detection limits suggest that with MIRC we are able to recover most planets $>2$ M$_J$ at orbits $>0.75$ AU around single components of intermediate mass close binary systems.

The Gaia mission will also use the astrometry method for discovering giant exoplanets. While Gaia is expected to be extremely successful in recovering massive planets around low mass stars, companions with mass $M_P$ $<$ 10 $M_J$ around around A and B-type stars will likely remain undetectable by Gaia. A common criterion for detection of an undiscovered exoplanet with Gaia is 
\begin{equation}
S/N = a (\frac{\sigma_\Lambda}{\sqrt{N}})^{-1} > 20, 
\end{equation}
where $\sigma_\Lambda$ is the single-epoch measurement error, $a$ is the semi-major axis of the detected orbit, and $N$ is the number of observations \citep{sahlmann2016}. Using this criteria for a discovery with $\sigma_\Lambda$ = 50 $\mu$-as and $N$ = 70 measurements over 5 years, a 1 M$_J$ planet on a 3 AU orbit around an A-type star of 2 M$_{\odot}$ could be detected out to 10 pc. Since there are just 4 A-type stars within 10 pc, Jupiter-mass planet discoveries around massive stars are expected to be rare with Gaia. A 10 M$_J$ planet on a 3 AU orbit around an A-star is detectable out to 100 pc, where there are over 400 A-type stars available for study \citep{derosa2014}. Thus, companions $\sim$10 M$_J$ and greater around A-type stars should be detectable with Gaia. With better single-epoch measurements, we plan to search for Jupiter-mass planets on orbits $<$5 AU around A and B-type stars which will complement the more massive companions discovered by Gaia.

\section{Summary}
Obtaining both spectroscopic and visual orbits of binary stars allows one to measure the full 3D orbit, masses, and parallax of the system. This information is crucial for testing models of stellar evolution. In this work we have obtained a highly precise visual orbit with $>2$ years of data from the MIRC instrument on the CHARA long-baseline interferometer. We also use 97 new spectra from Fairborn Observatory along with 87 unpublished spectra obtained at Lick Observatory by \citet{Duncan1979} to obtain a double-lined spectroscopic binary orbit. In our full binary analysis of $\delta$ Del we determine component masses of $1.78 \pm 0.07$ $M_{\odot}$ and $1.62 \pm 0.07$ $M_{\odot}$. We measure a distance of $63.61 \pm 0.89$ ($\pm 0.16$ systematic error)  pc, which differs from the revised Hipparcos value of $68.45 \pm 0.94$ pc. 

We find that the evolutionary state of $\delta$ Del is puzzling. Combining our H-band MIRC observations with R-band data from the PAVO instrument on CHARA, we are able to determine individual magnitudes and temperatures for each component. A metallicity of [Fe/H]=$-$0.5 is required to match our mass determination to MIST stellar models. The position on the HR diagram, however, implies that one component is more evolved than the other by $\sim$ 200 Myrs. We propose four possibilities for explaining this seemingly impossible evolutionary state: 1) stellar models are incorrect on the subgiant branch, 2) $\delta$ Scuti variables evolve differently than normal stars just after the main sequence, 3) interactions with a now-ejected tertiary companion created different mixing processes for each component or 4) the more massive component of $\delta$ Del is the result of a merger event at an age of $\sim$ 200 Myr which then evolved as a normal star. 

Because of the high precision of our visual orbit of $\delta$ Del, we calculate exoplanet detection limits around one of the two stars of
this binary system after accounting for the orbital motion of the
companion. With the MIRC instrument we have maintained $<10$ $\mu$-as precision on differential position over $>2$ years. This is the precision needed to detect Jupiter-mass planets at orbits up to a few AU. Though the presence of a planet around a component of $\delta$ Del is unlikely because of the extremely close binary separation, we have shown that if this precision can be extended to wider binaries MIRC is within reach of detecting planets $>2$ M$_J$ at orbits $>0.75$ AU. Developing this capability will allow us to search for exoplanets in regimes that are difficult to probe with RV and transit surveys, such as around hot binary stars. Our group is starting project ARMADA (ARrangement for Micro-Arcsecond Differential Astrometry), which will use MIRC at the CHARA array to target hot binary stars with the goal of detecting massive exoplanets on orbits up to a few AU around intermediate mass stars. 

\acknowledgments
This work is based upon observations obtained with the Georgia State
University Center for High Angular Resolution Astronomy Array at Mount
Wilson Observatory.  The CHARA Array is supported by the National
Science Foundation under Grants No. AST-1211929 and AST-1411654.
Institutional support has been provided from the GSU College of Arts
and Sciences and the GSU Office of the Vice President for Research and
Economic Development. This research has made use of the Jean-Marie Mariotti Center SearchCal service\footnote{available at \texttt{http://www.jmmc.fr/searchcal\_page.htm}}.  JDM and TG wish to gratefully acknowledge support by NASA XRP Grant NNX16AD43G. 
Astronomy at Tennessee State University is supported by the state of Tennessee through
its Centers of Excellence program. SK acknowledges support from an European Research Council Starting Grant (Grant Agreement No.\ 639889) and STFC Rutherford Fellowship (ST/J004030/1). D.H. acknowledges support by the National Aeronautics and Space Administration under Grant NNX14AB92G issued through the Kepler Participating Scientist Program. TRW acknowledges the support of the Villum Foundation (research grant 10118). We thank the anonymous referee for insightful comments and suggestions.

\vspace{5mm}
\facilities{CHARA, Fairborn Observatory, Lick Observatory}
\software{emcee, lmfit, astropy}

\bibliographystyle{aasjournal}
\bibliography{references}

\begin{thebibliography}{}
\expandafter\ifx\csname natexlab\endcsname\relax\def\natexlab#1{#1}\fi

\bibitem[{{Abt} \& {Morrell}(1995)}]{abt1995}
{Abt}, H.~A., \& {Morrell}, N.~I. 1995, \apjs, 99, 135

\bibitem[{{Andrievsky}(1997)}]{andrievsky1997}
{Andrievsky}, S.~M. 1997, \aap, 321, 838

\bibitem[{{Astropy Collaboration} {et~al.}(2013){Astropy Collaboration},
  {Robitaille}, {Tollerud}, {Greenfield}, {Droettboom}, {Bray}, {Aldcroft},
  {Davis}, {Ginsburg}, {Price-Whelan}, {Kerzendorf}, {Conley}, {Crighton},
  {Barbary}, {Muna}, {Ferguson}, {Grollier}, {Parikh}, {Nair}, {Unther},
  {Deil}, {Woillez}, {Conseil}, {Kramer}, {Turner}, {Singer}, {Fox}, {Weaver},
  {Zabalza}, {Edwards}, {Azalee Bostroem}, {Burke}, {Casey}, {Crawford},
  {Dencheva}, {Ely}, {Jenness}, {Labrie}, {Lim}, {Pierfederici}, {Pontzen},
  {Ptak}, {Refsdal}, {Servillat}, \& {Streicher}}]{astropy}
{Astropy Collaboration}, {Robitaille}, T.~P., {Tollerud}, E.~J., {et~al.} 2013,
  \aap, 558, A33

\bibitem[{{Baglin} {et~al.}(1973){Baglin}, {Breger}, {Chevalier}, {Hauck}, {Le
  Contel}, {Sareyan}, \& {Valtier}}]{baglin1973}
{Baglin}, A., {Breger}, M., {Chevalier}, C., {et~al.} 1973, \aap, 23, 221

\bibitem[{{Barnes} {et~al.}(1978){Barnes}, {Evans}, \& {Moffett}}]{barnes1978}
{Barnes}, T.~G., {Evans}, D.~S., \& {Moffett}, T.~J. 1978, \mnras, 183, 285

\bibitem[{{Bonneau} {et~al.}(2014){Bonneau}, {Millour}, \&
  {Meilland}}]{bonneau2014}
{Bonneau}, D., {Millour}, F., \& {Meilland}, A. 2014, in EAS Publications
  Series, Vol.~69, EAS Publications Series, 335--372

\bibitem[{{Breger}(1979)}]{breger1979}
{Breger}, M. 1979, \pasp, 91, 5

\bibitem[{{Breger}(1980)}]{breger1980}
---. 1980, \ssr, 27, 361

\bibitem[{{Casertano} {et~al.}(2008){Casertano}, {Lattanzi}, {Sozzetti},
  {Spagna}, {Jancart}, {Morbidelli}, {Pannunzio}, {Pourbaix}, \&
  {Queloz}}]{casertano2008}
{Casertano}, S., {Lattanzi}, M.~G., {Sozzetti}, A., {et~al.} 2008, \aap, 482,
  699

\bibitem[{{Castelli} \& {Kurucz}(2004)}]{castelli2004}
{Castelli}, F., \& {Kurucz}, R.~L. 2004, ArXiv Astrophysics e-prints,
  astro-ph/0405087

\bibitem[{{Cenarro} {et~al.}(2007){Cenarro}, {Peletier},
  {S{\'a}nchez-Bl{\'a}zquez}, {Selam}, {Toloba}, {Cardiel},
  {Falc{\'o}n-Barroso}, {Gorgas}, {Jim{\'e}nez-Vicente}, \&
  {Vazdekis}}]{cenarro2007}
{Cenarro}, A.~J., {Peletier}, R.~F., {S{\'a}nchez-Bl{\'a}zquez}, P., {et~al.}
  2007, \mnras, 374, 664

\bibitem[{{Chelli} {et~al.}(2016){Chelli}, {Duvert}, {Bourg{\`e}s}, {Mella},
  {Lafrasse}, {Bonneau}, \& {Chesneau}}]{chelli2016}
{Chelli}, A., {Duvert}, G., {Bourg{\`e}s}, L., {et~al.} 2016, \aap, 589, A112

\bibitem[{{Choi} {et~al.}(2016){Choi}, {Dotter}, {Conroy}, {Cantiello},
  {Paxton}, \& {Johnson}}]{mist2}
{Choi}, J., {Dotter}, A., {Conroy}, C., {et~al.} 2016, \apj, 823, 102

\bibitem[{{Cutri} {et~al.}(2003){Cutri}, {Skrutskie}, {van Dyk}, {Beichman},
  {Carpenter}, {Chester}, {Cambresy}, {Evans}, {Fowler}, {Gizis}, {Howard},
  {Huchra}, {Jarrett}, {Kopan}, {Kirkpatrick}, {Light}, {Marsh}, {McCallon},
  {Schneider}, {Stiening}, {Sykes}, {Weinberg}, {Wheaton}, {Wheelock}, \&
  {Zacarias}}]{2mass}
{Cutri}, R.~M., {Skrutskie}, M.~F., {van Dyk}, S., {et~al.} 2003, VizieR Online
  Data Catalog, 2246

\bibitem[{{de Mink} {et~al.}(2014){de Mink}, {Sana}, {Langer}, {Izzard}, \&
  {Schneider}}]{mink2014}
{de Mink}, S.~E., {Sana}, H., {Langer}, N., {Izzard}, R.~G., \& {Schneider},
  F.~R.~N. 2014, \apj, 782, 7

\bibitem[{{De Rosa} {et~al.}(2014){De Rosa}, {Patience}, {Wilson}, {Schneider},
  {Wiktorowicz}, {Vigan}, {Marois}, {Song}, {Macintosh}, {Graham}, {Doyon},
  {Bessell}, {Thomas}, \& {Lai}}]{derosa2014}
{De Rosa}, R.~J., {Patience}, J., {Wilson}, P.~A., {et~al.} 2014, \mnras, 437,
  1216

\bibitem[{{Dotter}(2016)}]{mist1}
{Dotter}, A. 2016, \apjs, 222, 8

\bibitem[{{Duncan} \& {Preston}(1979)}]{Duncan1979}
{Duncan}, D.~K., \& {Preston}, G.~W. 1979, in \baas, Vol.~11, Bulletin of the
  American Astronomical Society, 728

\bibitem[{{Eaton} \& {Williamson}(2007)}]{ew07}
{Eaton}, J.~A., \& {Williamson}, M.~H. 2007, \pasp, 119, 886

\bibitem[{{Eggen}(1956)}]{Eggen1956}
{Eggen}, O.~J. 1956, \pasp, 68, 541

\bibitem[{{Fekel} \& {Griffin}(2011)}]{fg11}
{Fekel}, F.~C., \& {Griffin}, R.~F. 2011, The Observatory, 131, 283

\bibitem[{{Fekel} {et~al.}(2013){Fekel}, {Rajabi}, {Muterspaugh}, \&
  {Williamson}}]{fetal13}
{Fekel}, F.~C., {Rajabi}, S., {Muterspaugh}, M.~W., \& {Williamson}, M.~H.
  2013, \aj, 145, 111

\bibitem[{{Fekel} {et~al.}(2009){Fekel}, {Tomkin}, \& {Williamson}}]{ftw09}
{Fekel}, F.~C., {Tomkin}, J., \& {Williamson}, M.~H. 2009, \aj, 137, 3900

\bibitem[{{Feng} {et~al.}(2016){Feng}, {Tuomi}, {Jones}, {Butler}, \&
  {Vogt}}]{Feng2016}
{Feng}, F., {Tuomi}, M., {Jones}, H.~R.~A., {Butler}, R.~P., \& {Vogt}, S.
  2016, \mnras, 461, 2440

\bibitem[{{Foreman-Mackey} {et~al.}(2013){Foreman-Mackey}, {Hogg}, {Lang}, \&
  {Goodman}}]{Foreman2013}
{Foreman-Mackey}, D., {Hogg}, D.~W., {Lang}, D., \& {Goodman}, J. 2013, \pasp,
  125, 306

\bibitem[{{Gelman} \& {Rubin}(1992)}]{gelman1992}
{Gelman}, A., \& {Rubin}, D.~B. 1992, Statistical Science, 7, 457

\bibitem[{{Gray} \& {Garrison}(1989)}]{gray1989}
{Gray}, R.~O., \& {Garrison}, R.~F. 1989, \apjs, 69, 301

\bibitem[{{Gray} {et~al.}(2001){Gray}, {Napier}, \& {Winkler}}]{gray2001}
{Gray}, R.~O., {Napier}, M.~G., \& {Winkler}, L.~I. 2001, \aj, 121, 2148

\bibitem[{{Guzik} {et~al.}(1998){Guzik}, {Templeton}, \& {Bradley}}]{guzik1998}
{Guzik}, J.~A., {Templeton}, M.~R., \& {Bradley}, P.~A. 1998, in Astronomical
  Society of the Pacific Conference Series, Vol. 135, A Half Century of Stellar
  Pulsation Interpretation, ed. P.~A. {Bradley} \& J.~A. {Guzik}, 470

\bibitem[{{Hut}(1981)}]{hut1981}
{Hut}, P. 1981, \aap, 99, 126

\bibitem[{{Ireland} {et~al.}(2008){Ireland}, {M{\'e}rand}, {ten Brummelaar},
  {Tuthill}, {Schaefer}, {Turner}, {Sturmann}, {Sturmann}, \&
  {McAlister}}]{ireland2008}
{Ireland}, M.~J., {M{\'e}rand}, A., {ten Brummelaar}, T.~A., {et~al.} 2008, in
  \procspie, Vol. 7013, Optical and Infrared Interferometry, 701324

\bibitem[{{Kervella} {et~al.}(2004){Kervella}, {Th{\'e}venin}, {Di Folco}, \&
  {S{\'e}gransan}}]{kervella2004}
{Kervella}, P., {Th{\'e}venin}, F., {Di Folco}, E., \& {S{\'e}gransan}, D.
  2004, \aap, 426, 297

\bibitem[{{Liddle}(2007)}]{liddle2007}
{Liddle}, A.~R. 2007, \mnras, 377, L74

\bibitem[{{Lucy}(2014)}]{Lucy2014}
{Lucy}, L.~B. 2014, \aap, 563, A126

\bibitem[{{Matthews} \& {Mathieu}(1992)}]{matthews1992}
{Matthews}, L.~D., \& {Mathieu}, R.~D. 1992, in Astronomical Society of the
  Pacific Conference Series, Vol.~32, IAU Colloq. 135: Complementary Approaches
  to Double and Multiple Star Research, ed. H.~A. {McAlister} \& W.~I.
  {Hartkopf}, 244

\bibitem[{{Monnier} {et~al.}(2006){Monnier}, {Pedretti}, {Thureau}, {Berger},
  {Millan-Gabet}, {ten Brummelaar}, {McAlister}, {Sturmann}, {Sturmann},
  {Muirhead}, {Tannirkulam}, {Webster}, \& {Zhao}}]{Monnier2006}
{Monnier}, J.~D., {Pedretti}, E., {Thureau}, N., {et~al.} 2006, in \procspie,
  Vol. 6268, Society of Photo-Optical Instrumentation Engineers (SPIE)
  Conference Series, 62681P

\bibitem[{{Monnier} {et~al.}(2012){Monnier}, {Che}, {Zhao}, {Ekstr{\"o}m},
  {Maestro}, {Aufdenberg}, {Baron}, {Georgy}, {Kraus}, {McAlister}, {Pedretti},
  {Ridgway}, {Sturmann}, {Sturmann}, {ten Brummelaar}, {Thureau}, {Turner}, \&
  {Tuthill}}]{monnier2012}
{Monnier}, J.~D., {Che}, X., {Zhao}, M., {et~al.} 2012, \apjl, 761, L3

\bibitem[{{Morel} \& {Magnenat}(1978)}]{morel1978}
{Morel}, M., \& {Magnenat}, P. 1978, \aaps, 34, 477

\bibitem[{{Morgan} \& {Abt}(1972)}]{morgan1972}
{Morgan}, W.~W., \& {Abt}, H.~A. 1972, \aj, 77, 35

\bibitem[{{Motalebi} {et~al.}(2015){Motalebi}, {Udry}, {Gillon}, {Lovis},
  {S{\'e}gransan}, {Buchhave}, {Demory}, {Malavolta}, {Dressing}, {Sasselov},
  {Rice}, {Charbonneau}, {Collier Cameron}, {Latham}, {Molinari}, {Pepe},
  {Affer}, {Bonomo}, {Cosentino}, {Dumusque}, {Figueira}, {Fiorenzano},
  {Gettel}, {Harutyunyan}, {Haywood}, {Johnson}, {Lopez}, {Lopez-Morales},
  {Mayor}, {Micela}, {Mortier}, {Nascimbeni}, {Philips}, {Piotto}, {Pollacco},
  {Queloz}, {Sozzetti}, {Vanderburg}, \& {Watson}}]{Motalebi2015}
{Motalebi}, F., {Udry}, S., {Gillon}, M., {et~al.} 2015, \aap, 584, A72

\bibitem[{{Mourard} {et~al.}(2015){Mourard}, {Monnier}, {Meilland}, {Gies},
  {Millour}, {Benisty}, {Che}, {Grundstrom}, {Ligi}, {Schaefer}, {Baron},
  {Kraus}, {Zhao}, {Pedretti}, {Berio}, {Clausse}, {Nardetto}, {Perraut},
  {Spang}, {Stee}, {Tallon-Bosc}, {McAlister}, {ten Brummelaar}, {Ridgway},
  {Sturmann}, {Sturmann}, {Turner}, \& {Farrington}}]{mourard2015}
{Mourard}, D., {Monnier}, J.~D., {Meilland}, A., {et~al.} 2015, \aap, 577, A51

\bibitem[{{Murdoch} {et~al.}(1993){Murdoch}, {Hearnshaw}, \&
  {Clark}}]{murdoch1993}
{Murdoch}, K.~A., {Hearnshaw}, J.~B., \& {Clark}, M. 1993, \apj, 413, 349

\bibitem[{{Muterspaugh} {et~al.}(2010){Muterspaugh}, {Lane}, {Kulkarni},
  {Konacki}, {Burke}, {Colavita}, {Shao}, {Hartkopf}, {Boss}, \&
  {Williamson}}]{Muterspaugh2010}
{Muterspaugh}, M.~W., {Lane}, B.~F., {Kulkarni}, S.~R., {et~al.} 2010, \aj,
  140, 1657

\bibitem[{Newville {et~al.}(2014)Newville, Stensitzki, Allen, \&
  Ingargiola}]{Newville2014}
Newville, M., Stensitzki, T., Allen, D.~B., \& Ingargiola, A. 2014, {LMFIT:
  Non-Linear Least-Square Minimization and Curve-Fitting for Python¶}, , ,
  doi:10.5281/zenodo.11813

\bibitem[{{Niu} {et~al.}(2017){Niu}, {Fu}, {Li}, {Yang}, {Zong}, {Xue},
  {Zhang}, {Liu}, {Du}, \& {Zuo}}]{niu2017}
{Niu}, J.-S., {Fu}, J.-N., {Li}, Y., {et~al.} 2017, \mnras, 467, 3122

\bibitem[{{North} {et~al.}(1997){North}, {Jaschek}, \& {Egret}}]{north1997}
{North}, P., {Jaschek}, C., \& {Egret}, D. 1997, in ESA Special Publication,
  Vol. 402, Hipparcos - Venice '97, ed. R.~M. {Bonnet}, E.~{H{\o}g}, P.~L.
  {Bernacca}, L.~{Emiliani}, A.~{Blaauw}, C.~{Turon}, J.~{Kovalevsky},
  L.~{Lindegren}, H.~{Hassan}, M.~{Bouffard}, B.~{Strim}, D.~{Heger}, M.~A.~C.
  {Perryman}, \& L.~{Woltjer}, 367--370

\bibitem[{{Pauls} {et~al.}(2005){Pauls}, {Young}, {Cotton}, \&
  {Monnier}}]{pauls2005}
{Pauls}, T.~A., {Young}, J.~S., {Cotton}, W.~D., \& {Monnier}, J.~D. 2005,
  \pasp, 117, 1255

\bibitem[{{Paxton} {et~al.}(2011){Paxton}, {Bildsten}, {Dotter}, {Herwig},
  {Lesaffre}, \& {Timmes}}]{mist3}
{Paxton}, B., {Bildsten}, L., {Dotter}, A., {et~al.} 2011, \apjs, 192, 3

\bibitem[{{Paxton} {et~al.}(2013){Paxton}, {Cantiello}, {Arras}, {Bildsten},
  {Brown}, {Dotter}, {Mankovich}, {Montgomery}, {Stello}, {Timmes}, \&
  {Townsend}}]{mist4}
{Paxton}, B., {Cantiello}, M., {Arras}, P., {et~al.} 2013, \apjs, 208, 4

\bibitem[{{Paxton} {et~al.}(2015){Paxton}, {Marchant}, {Schwab}, {Bauer},
  {Bildsten}, {Cantiello}, {Dessart}, {Farmer}, {Hu}, {Langer}, {Townsend},
  {Townsley}, \& {Timmes}}]{mist5}
{Paxton}, B., {Marchant}, P., {Schwab}, J., {et~al.} 2015, \apjs, 220, 15

\bibitem[{{Perryman} {et~al.}(2014){Perryman}, {Hartman}, {Bakos}, \&
  {Lindegren}}]{Perryman2014}
{Perryman}, M., {Hartman}, J., {Bakos}, G.~{\'A}., \& {Lindegren}, L. 2014,
  \apj, 797, 14

\bibitem[{{Perryman} {et~al.}(1997){Perryman}, {Lindegren}, {Kovalevsky},
  {Hoeg}, {Bastian}, {Bernacca}, {Cr{\'e}z{\'e}}, {Donati}, {Grenon},
  {Grewing}, {van Leeuwen}, {van der Marel}, {Mignard}, {Murray}, {Le Poole},
  {Schrijver}, {Turon}, {Arenou}, {Froeschl{\'e}}, \&
  {Petersen}}]{Perryman1997}
{Perryman}, M.~A.~C., {Lindegren}, L., {Kovalevsky}, J., {et~al.} 1997, \aap,
  323, L49

\bibitem[{{Petersen} \& {Christensen-Dalsgaard}(1996)}]{peterson1996}
{Petersen}, J.~O., \& {Christensen-Dalsgaard}, J. 1996, \aap, 312, 463

\bibitem[{{Ranalli} {et~al.}(2017){Ranalli}, {Hobbs}, \&
  {Lindegren}}]{Ranalli2017}
{Ranalli}, P., {Hobbs}, D., \& {Lindegren}, L. 2017, ArXiv e-prints,
  arXiv:1704.02493

\bibitem[{{Reimers}(1976)}]{reimers1976}
{Reimers}, D. 1976, \aap, 53, 377

\bibitem[{{Sahlmann} {et~al.}(2014){Sahlmann}, {Lazorenko}, {S{\'e}gransan},
  {Mart{\'{\i}}n}, {Mayor}, {Queloz}, \& {Udry}}]{sahlmann2014}
{Sahlmann}, J., {Lazorenko}, P.~F., {S{\'e}gransan}, D., {et~al.} 2014,
  \memsai, 85, 674

\bibitem[{{Sahlmann} {et~al.}(2016){Sahlmann}, {Mart{\'{\i}}n-Fleitas}, {Mora},
  {Abreu}, {Crowley}, \& {Joliet}}]{sahlmann2016}
{Sahlmann}, J., {Mart{\'{\i}}n-Fleitas}, J., {Mora}, A., {et~al.} 2016, in
  \procspie, Vol. 9904, Space Telescopes and Instrumentation 2016: Optical,
  Infrared, and Millimeter Wave, 99042E

\bibitem[{{Sandberg Lacy} \& {Fekel}(2011)}]{lf11}
{Sandberg Lacy}, C.~H., \& {Fekel}, F.~C. 2011, \aj, 142, 185

\bibitem[{{Sato} {et~al.}(2015){Sato}, {Hirano}, {Omiya}, {Harakawa},
  {Kobayashi}, {Hasegawa}, {Takarada}, {Kawauchi}, \& {Masuda}}]{Sato2015}
{Sato}, B., {Hirano}, T., {Omiya}, M., {et~al.} 2015, \apj, 802, 57

\bibitem[{{Scarfe}(2010)}]{s10}
{Scarfe}, C.~D. 2010, The Observatory, 130, 214

\bibitem[{{Sozzetti} {et~al.}(2014){Sozzetti}, {Giacobbe}, {Lattanzi},
  {Micela}, {Morbidelli}, \& {Tinetti}}]{sozzetti2014}
{Sozzetti}, A., {Giacobbe}, P., {Lattanzi}, M.~G., {et~al.} 2014, \mnras, 437,
  497

\bibitem[{{Struve} {et~al.}(1957){Struve}, {Sahade}, \& {Zebergs}}]{Struve1957}
{Struve}, O., {Sahade}, J., \& {Zebergs}, V. 1957, \apj, 125, 692

\bibitem[{{Tassoul}(1987)}]{tassoul1987}
{Tassoul}, J.-L. 1987, \apj, 322, 856

\bibitem[{{Tassoul} \& {Tassoul}(1992)}]{tassoul1992}
{Tassoul}, J.-L., \& {Tassoul}, M. 1992, \apj, 395, 259

\bibitem[{{ten Brummelaar} {et~al.}(2005){ten Brummelaar}, {McAlister},
  {Ridgway}, {Bagnuolo}, {Turner}, {Sturmann}, {Sturmann}, {Berger}, {Ogden},
  {Cadman}, {Hartkopf}, {Hopper}, \& {Shure}}]{tenbrummelaar2005}
{ten Brummelaar}, T.~A., {McAlister}, H.~A., {Ridgway}, S.~T., {et~al.} 2005,
  \apj, 628, 453

\bibitem[{{Tokovinin}(2004)}]{tokovinin2004}
{Tokovinin}, A. 2004, in Revista Mexicana de Astronomia y Astrofisica, vol.~27,
  Vol.~21, Revista Mexicana de Astronomia y Astrofisica Conference Series, ed.
  C.~{Allen} \& C.~{Scarfe}, 7--14

\bibitem[{{Torres} {et~al.}(2010){Torres}, {Andersen}, \&
  {Gim{\'e}nez}}]{torres2010}
{Torres}, G., {Andersen}, J., \& {Gim{\'e}nez}, A. 2010, \aapr, 18, 67

\bibitem[{{Tsvetkov}(1990)}]{tsvetkov1990}
{Tsvetkov}, T.~G. 1990, \apss, 173, 1

\bibitem[{{van Leeuwen}(2007)}]{vanLeeuwen2007}
{van Leeuwen}, F. 2007, \aap, 474, 653

\bibitem[{{White} {et~al.}(2013){White}, {Huber}, {Maestro}, {Bedding},
  {Ireland}, {Baron}, {Boyajian}, {Che}, {Monnier}, {Pope}, {Roettenbacher},
  {Stello}, {Tuthill}, {Farrington}, {Goldfinger}, {McAlister}, {Schaefer},
  {Sturmann}, {Sturmann}, {ten Brummelaar}, \& {Turner}}]{white2013}
{White}, T.~R., {Huber}, D., {Maestro}, V., {et~al.} 2013, \mnras, 433, 1262

\bibitem[{{Wright} \& {Howard}(2009)}]{Wright2009}
{Wright}, J.~T., \& {Howard}, A.~W. 2009, \apjs, 182, 205

\bibitem[{{Zahn}(1977)}]{zahn1977}
{Zahn}, J.-P. 1977, \aap, 57, 383

\end{thebibliography}

\listofchanges

\appendix
\section{$\delta$ Scuti Pulsations}
\label{appendix}
Because of short period variations in their radial velocity curves, both components of $\delta$ Del have been previously classified as $\delta$~Scuti variables with periods of $0.158 \pm 0.006$ days for the primary (more massive) component and $0.134 \pm 0.015$ days for the secondary \citep{Duncan1979}. However, this previous analysis of the 1979 Lick Observatory data also concluded that there are multiple periodicities in the $\delta$ Scuti pulsations. Hence, fitting the pulsations to a single sinusoid with the peak period does not capture the true nature of these variations. More evenly sampled data at all epochs is likely needed to model these pulsations thoroughly. Nevertheless, we describe a "first-order" correction of these pulsations in order to improve the overall RV fit. 

We first carry out a least-squares fit to all of the radial velocity data from Fairborn and Lick observatories, and we subtract out the resulting best-fit RV for each data point. We then search for additional periodic signals in the residuals by generating a Lomb-Scargle periodogram with the built-in function from the \textit{astropy} package \citep{astropy}. A single sinusoid is fit to the residual data with a period determined from the highest peak of the periodogram. The significance of a peak is determined by estimating the false-alarm probability (FAP) using the bootstrap method described in \citep{murdoch1993}. 

\begin{figure}
\centering
\begin{minipage}[b]{.45\textwidth}
\centering
\includegraphics[width=\linewidth]{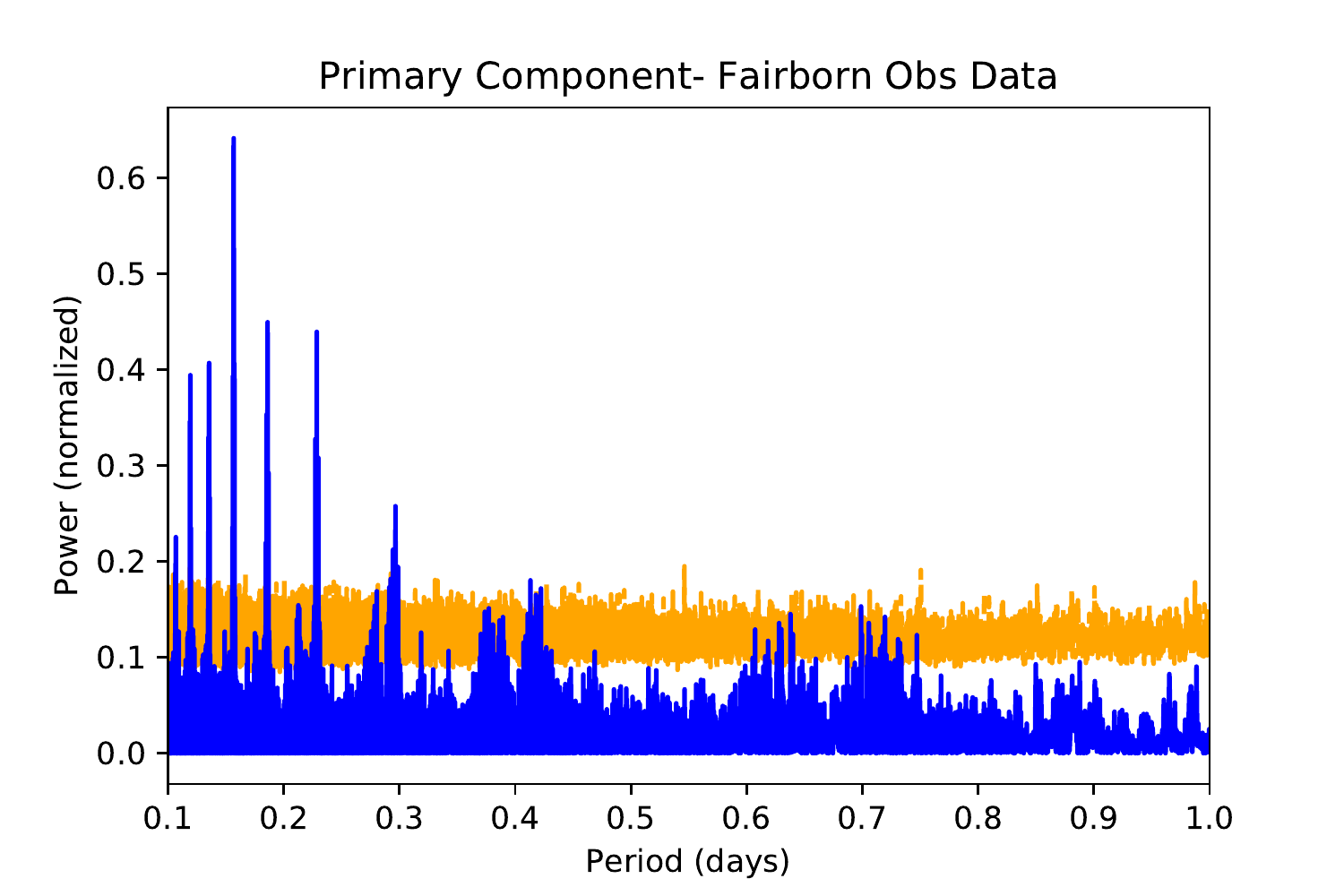}
\end{minipage}
\begin{minipage}[b]{.45\textwidth}
\centering
\includegraphics[width=\linewidth]{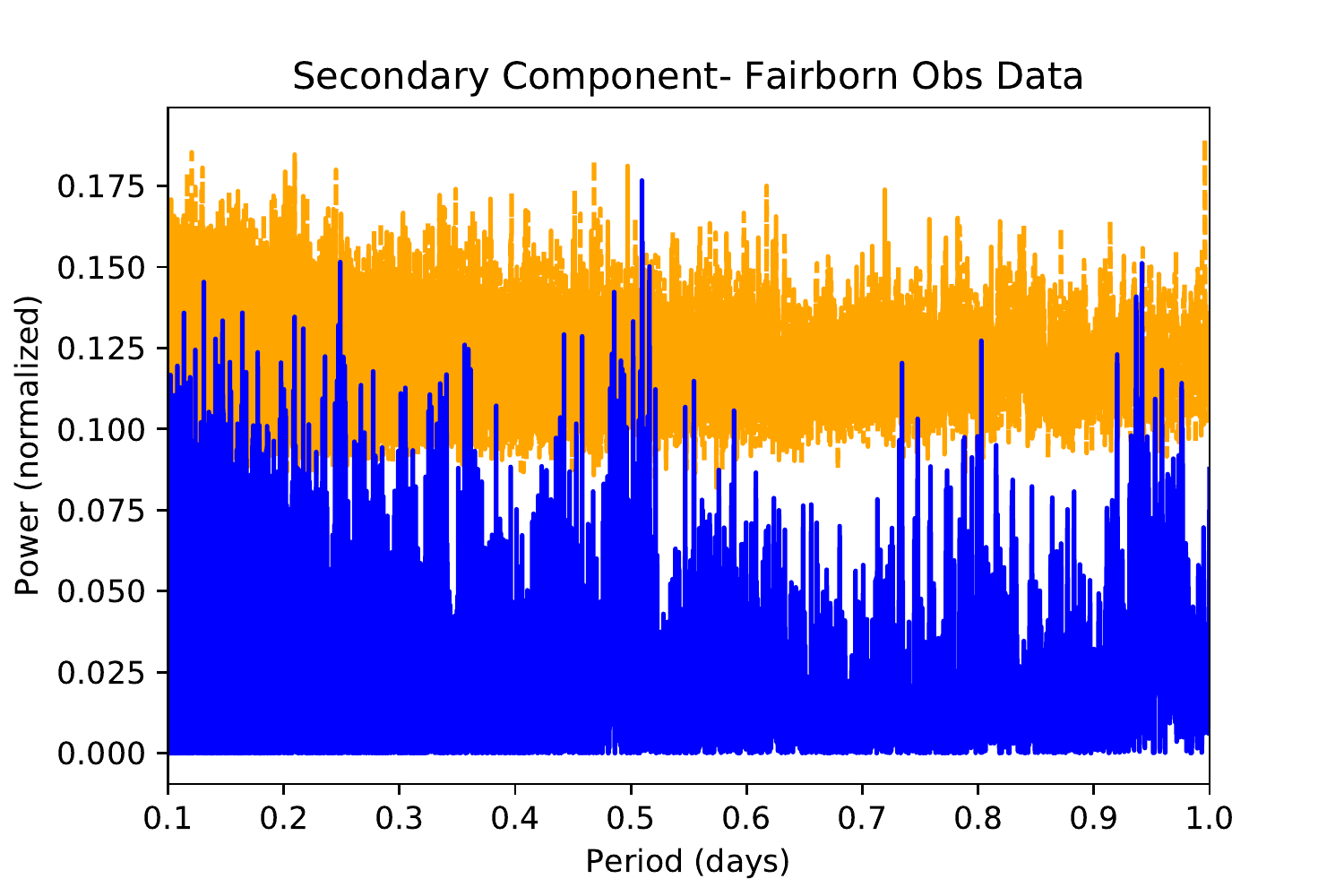}
\end{minipage}

\begin{minipage}[b]{.45\textwidth}
\centering
\includegraphics[width=\linewidth]{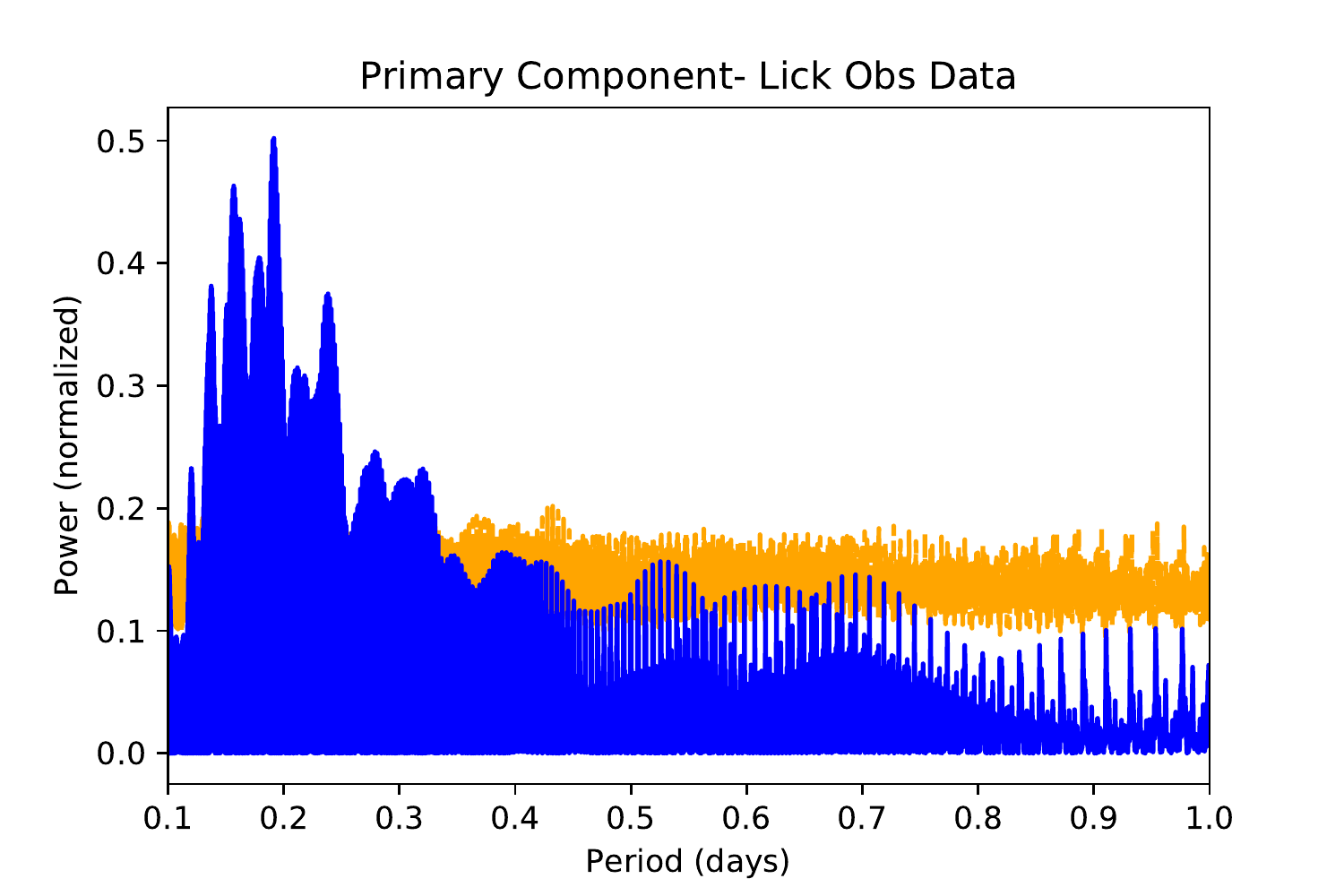}
\end{minipage}
\begin{minipage}[b]{.45\textwidth}
\centering
\includegraphics[width=\linewidth]{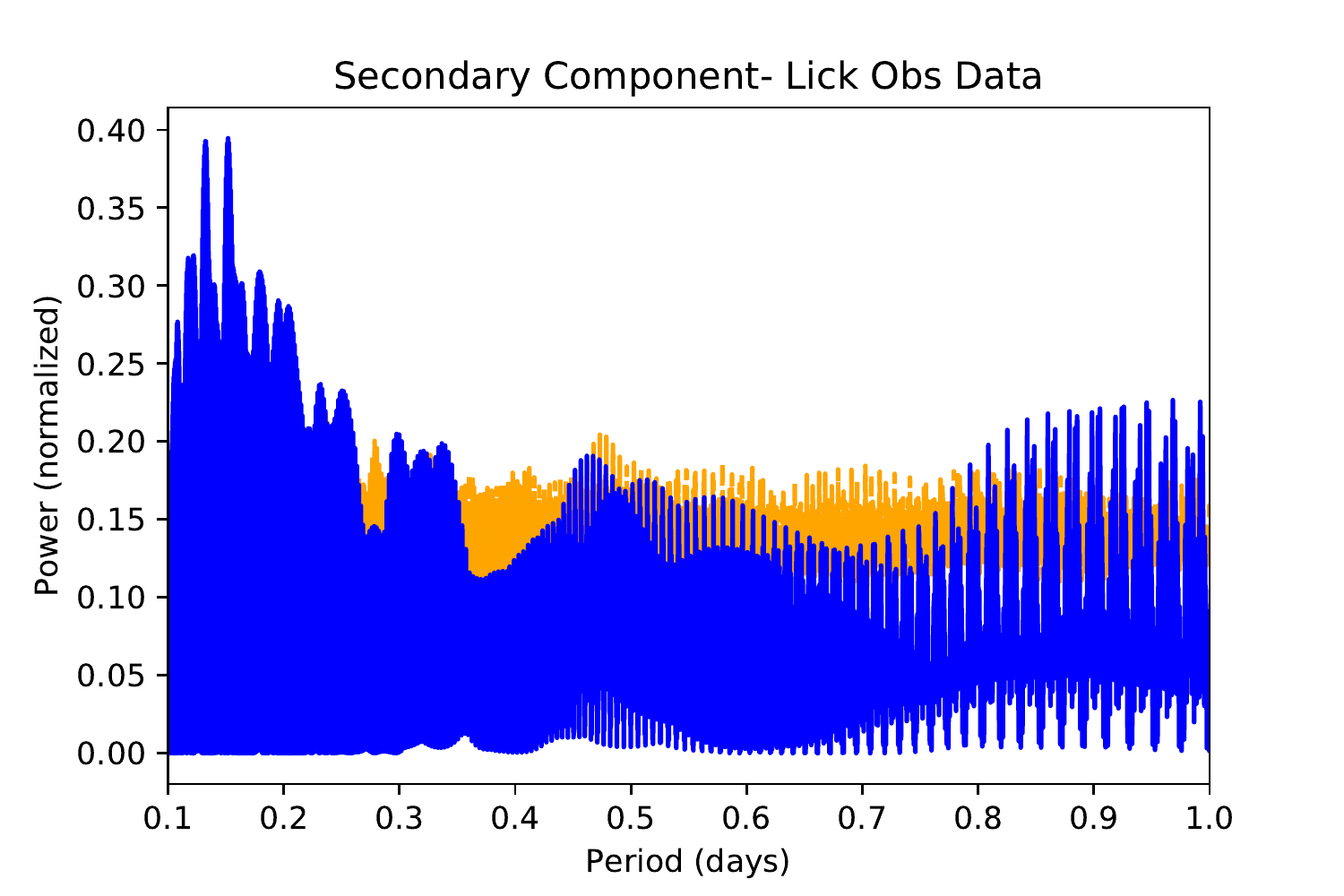}
\end{minipage}
\caption{After subtracting out the best-fit RV orbit, we search for additional periodicity in the data due to $\delta$ Scuti pulsations. We detect significant peaks in the primary component of the Fairborn data, and both components of the Lick data. The orange background signal depicts peaks of 3-sigma significance determined by bootstrapping. The peaks in the secondary component of the Fairborn data are not highly significant.}
\label{periodograms}
\end{figure}

\begin{figure}
\centering
\begin{minipage}[b]{.4\textwidth}
\centering
\includegraphics[width=\linewidth]{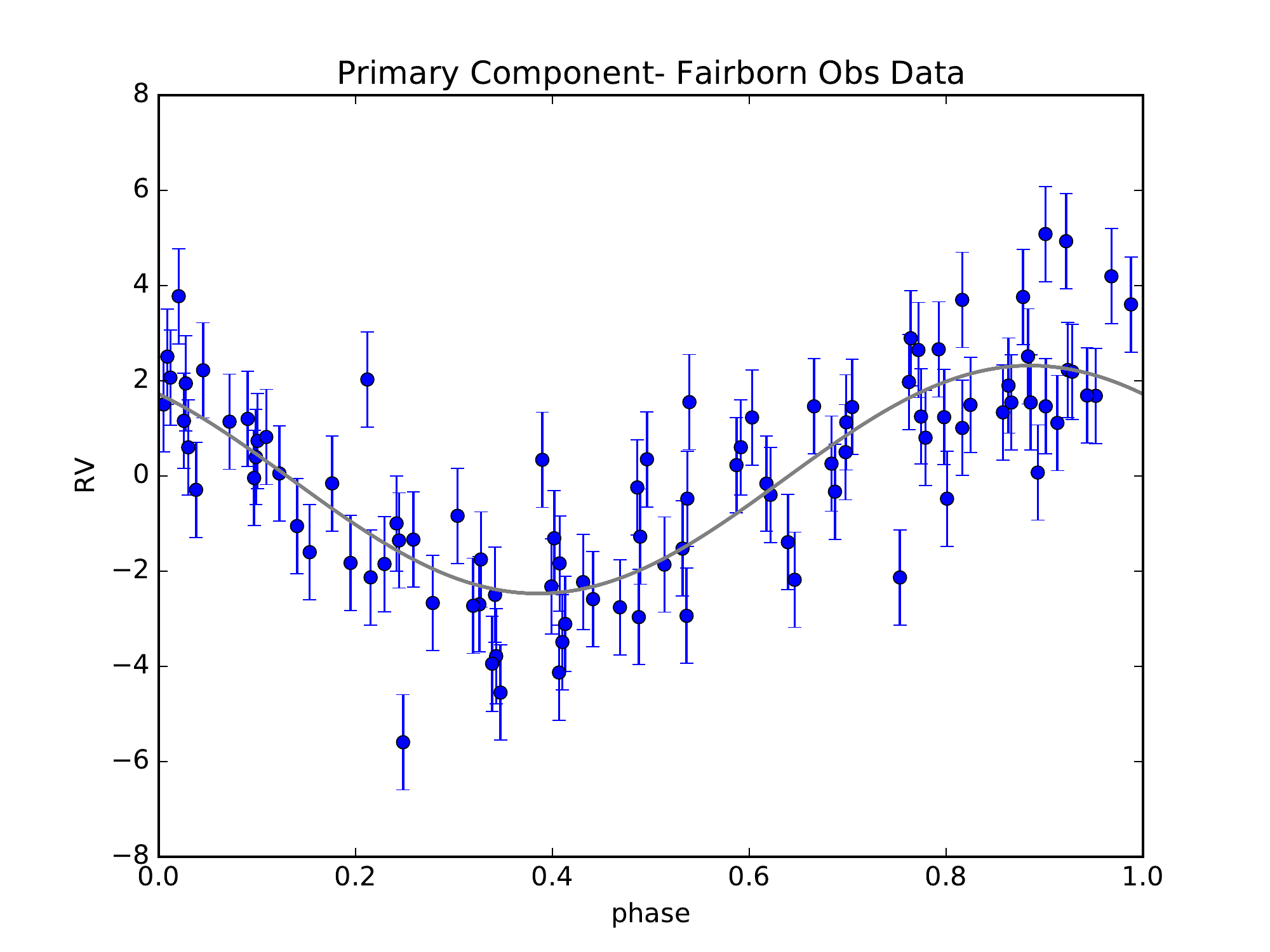}
\end{minipage}

\begin{minipage}[b]{.45\textwidth}
\centering
\includegraphics[width=\linewidth]{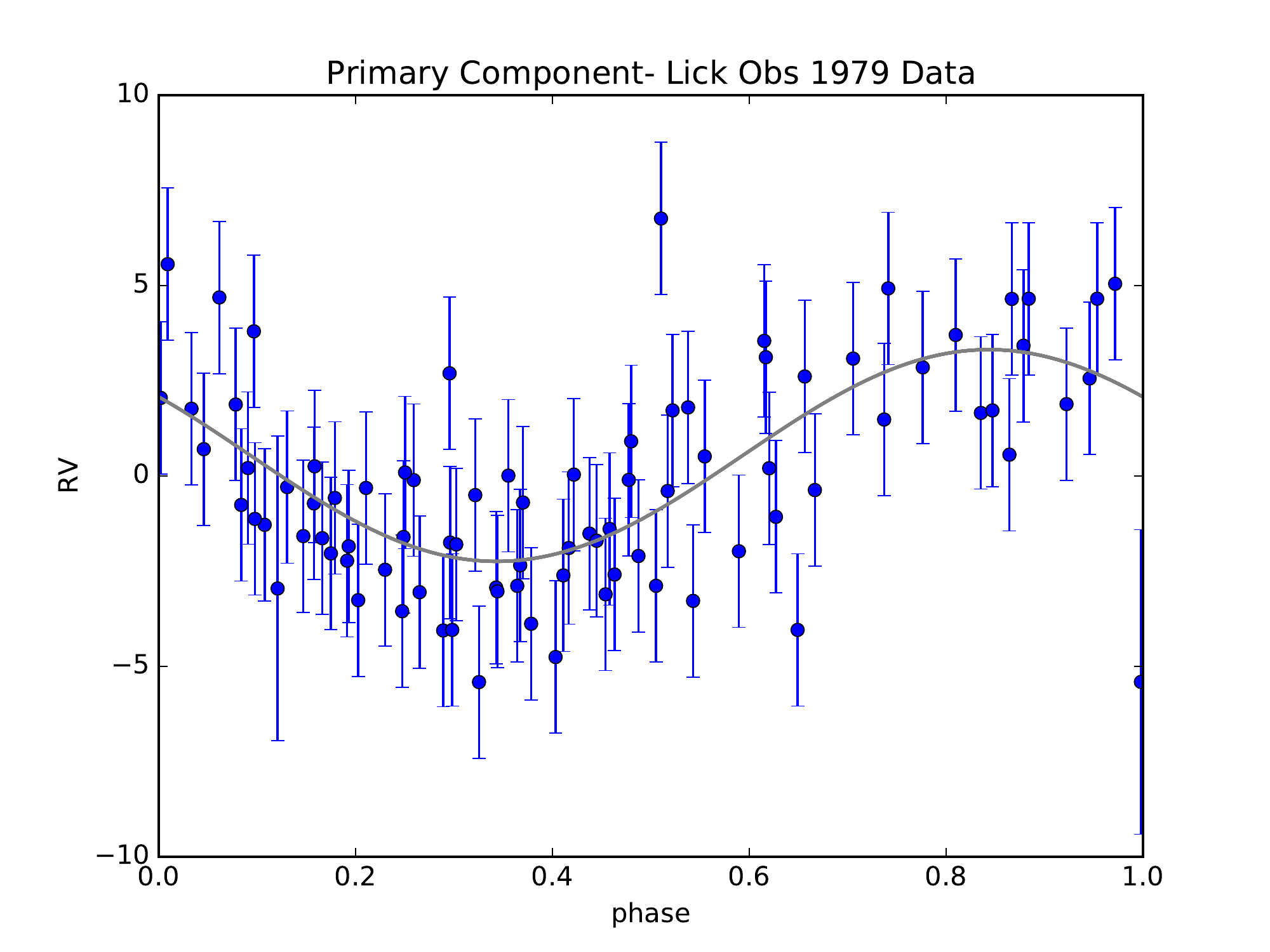}
\end{minipage}
\begin{minipage}[b]{.45\textwidth}
\centering
\includegraphics[width=\linewidth]{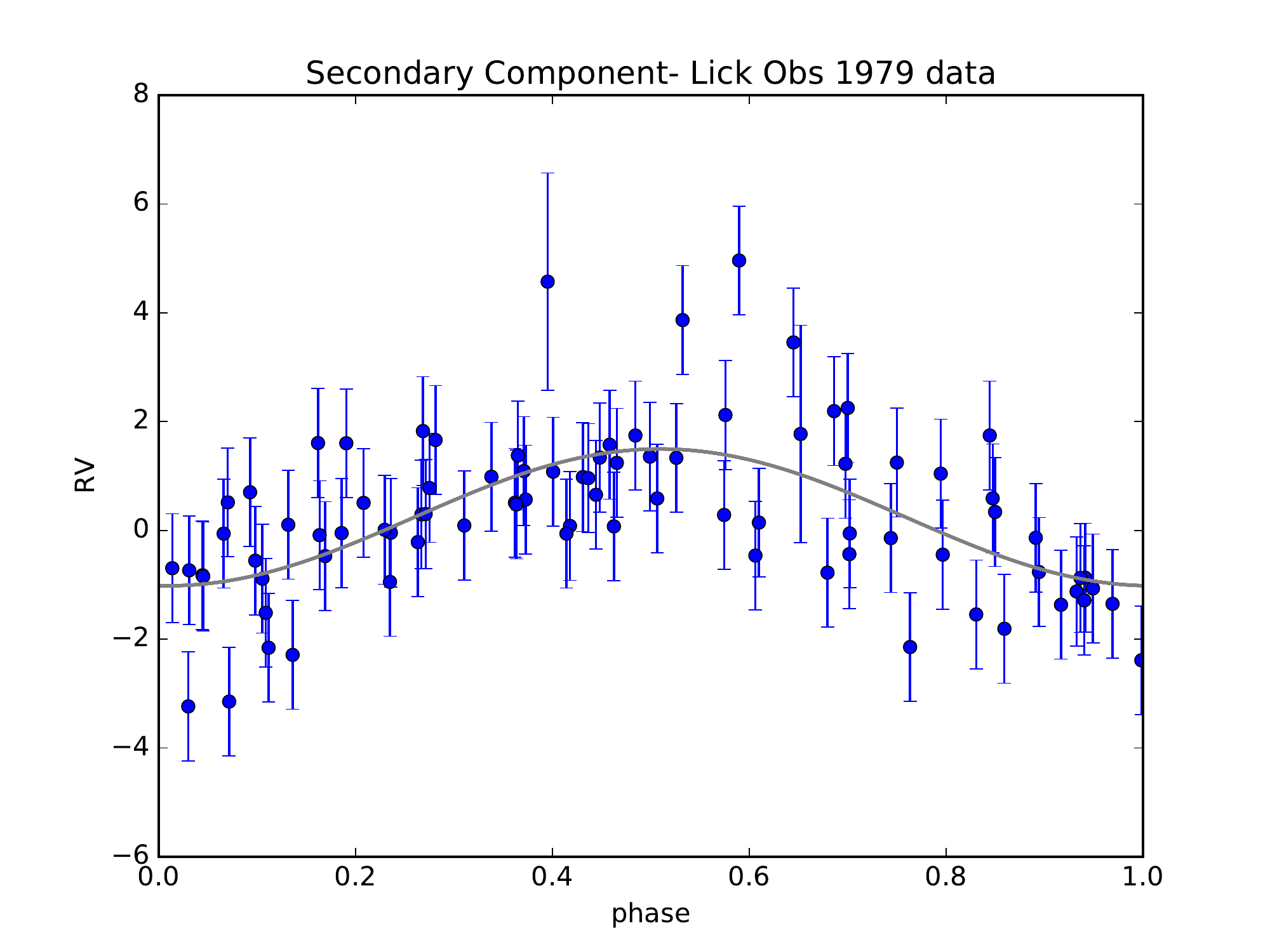}
\end{minipage}
\caption{$\delta$ Scuti pulsations in the residuals for the Fairborn and Lick RV data after the best fit orbit is subtracted out. The primary component of both datasets has a period of 0.157 days. The secondary in the Lick data has a period of 0.132 days. We detect no significant period signal for the secondary in the Fairborn Observatory data. Figures are phase-folded with $T_0=56823.6$.}
\label{scuti_fits}
\end{figure}

In the Fairborn Obs data, we find a significant peak in the primary at 0.157 days. The secondary component, however, shows no significant peaks in the Fairborn data. The strongest peak in the periodogram has a FAP of $\sim$0.91, suggesting that it is not a true signal. Thus we do not model any pulsations in this component for the Fairborn data. We detect significant peaks in both components of the 1979 Lick Observatory data, though the periodograms show peaks for many different periods. For the secondary component in the Lick Observatory data we model the pulsations with the first peak at 0.1323 days, within the error bars of the 1979 analysis. Since there is also a peak at 0.157 days for the primary component in the Lick data, we again use this period to model the pulsations of the primary. We subtract the $\delta$ Scuti pulsations out of the radial velocity data, separately for the Fairborn and Lick Observatory velocities, and re-fit the resulting data with our RV model. Our reduced $\chi^2$ value for the RV fit decreases from 3.5 to 1.8 after subtracting out the pulsations. The periodograms for each dataset are shown in Figure \ref{periodograms}. Figure \ref{scuti_fits} shows the $\delta$ Scuti pulsations of the primary in the Fairborn data and both components in the Lick data.

\end{document}